\documentclass{ccjnl}

\title{Sensing-assisted Accurate and Fast Beam Management for Cellular-connected mmWave UAV Network}
\author{Yanpeng Cui\inst{1}, Qixun Zhang\inst{1}, Zhiyong Feng\inst{1*}, Qin Wen\inst{2}, Ying Zhou\inst{1}, Zhiqing Wei\inst{1}, Ping Zhang\inst{1}\corinfo{fengzy@bupt.edu.cn}}
\receiveddate{2023.02.27}
\reviseddate{2023.05.11}
\Editor{}

\address[1]{School of Information and Communication Engineering, Beijing University of Posts and Telecommunications, Beijing 100876, China}
\address[2]{School of Computer Science, Northwestern Polytechnical University, Xi'an Shaanxi, 710072, China}

\begin{document}
\maketitle

\begin{abstract}
Beam management, including initial access (IA) and beam tracking, is essential to the millimeter-wave Unmanned Aerial Vehicle (UAV) network. However, the conventional communication-only and feedback-based schemes suffer a high delay and low accuracy of beam alignment, since they only enable the receiver passively ``hear'' the information of the transmitter from the radio domain. This paper presents a novel sensing-assisted beam management approach, the first solution that fully utilizes the information from the visual domain to improve communication performance. We employ both integrated sensing and communication and computer vision techniques and design an extended Kalman filtering method for beam tracking and prediction. Besides, we also propose a novel dual identity association solution to distinguish multiple UAVs in dynamic environments. Real-world experiments and numerical results show that the proposed solution outperforms the conventional methods in IA delay, association accuracy, tracking error, and communication performance.
\end{abstract}

\section{introduction}
\label{s1}

With the integration of wireless communications and aerial vehicular technology, we are at the dawn of the era of ubiquitous aerial networks, in which unmanned aerial vehicles (UAVs) are connected seamlessly to enable profound progress in numerous applications \cite{cui2022topology}. Under such emerging scenarios, the naive maximalist approach of blindly improving data transmission capabilities may no longer be pursued. Instead, the future UAV network paradigm may shift towards generating the most valuable information that should be instantly and efficiently transmitted at the right time \cite{on-purpose}. 

In the millimeter wave (mmWave) communication systems, narrow beams are utilized to compensate for high path loss and avoid interference to unintended receivers \cite{millimeter-wave-UAV-survey}. Nevertheless, along with the benefits come additional troubles. For instance, given that the discontinuous reception (DRX) mode \cite{DRX} is frequently utilized due to the limited energy of the UAVs, it will suffer serious delay issues when the network switches from idle mode (IM) to connected mode (CM), or keeps working in CM. These delay issues generally occur during the beam management process, including 1) initial access (IA), which allows the UAVs to establish a physical link connection with the BS, and 2) beam tracking, which keeps the alignment of the fine beams as well as enables beam adaptation, path selection, and radio link failure recovery procedures.

When the network is switching from IM to CM, the IA procedure is incredibly time-consuming since both the transmitter and receiver have no knowledge of the channel state. For instance, the primitive exhaustive search, which makes the transmitter and receiver consecutively check all the directions and select the one with the highest signal-to-noise ratio (SNR) \cite{standalone-magazine}, and suffers the highest delay. The hierarchical sequential search approach employs hierarchical beam codebooks to reduce the IA delay \cite{interactive-search}. However, it causes a disparity between the range of detection and service and still faces a long IA delay. The position information (PI)-based methods improve the efficiency by the prior distribution information of the receiver’s locations \cite{Position-aided-beam-learning}, which reduces the search space. However, the potential beam direction may differ much from the best one, and storing the beam for any position necessitates an exhaustive search once the IA failure occurs. In addition, the employment of non-standalone network architectures requires deploying the macro base station, which increases the cost \cite{dual-band-search}.

When the network keeps working in CM, it also suffers a time-consuming beam alignment. The conventional beam training methods require the transmitter to send some pilot to the receiver so that the latter could estimate the channel states and feed them back to the former. This will definitely lead to a large delay and may lose the fleeting alignment opportunity. The state-of-the-art beam tracking approaches have exploited the temporal correlation between consecutive transmission blocks to update the beam information, which needs only a small number of pilots and reduces the time consumption. Existing works on predictive beam tracking are generally based on the Kalman filtering method to meet the critical latency requirement. Nevertheless, these communication-only feedback schemes typically utilize only a small number of pilots, which causes a finite matched filtering gain for angle estimation and loss of localization accuracy. Besides, the estimations of the channel that feedback from the previous period may be outdated in the current period, which is not suitable for high-mobility UAV communication scenarios.

The above-mentioned solutions induce a bottleneck in improving efficiency for the following reasons. The conventional communication-only approaches only passively ``hear" the global unique temporary identity (GUTI) \cite{Dual_Identity} and physical characteristics of the receiver from the radio domain, which we vividly call the auditory domain (AD). The AD-only transmitter is more like a ``blind man" with the perfect auditory ability and only uses the AD information, resulting in unbearable delay and accuracy \cite{Dual_Identity}. In fact, the transmitter should have the capability to actively utilize the rich information from the ``visual domain (VD)" to accelerate the link establishment and ensure tracking accuracy \cite{Dual_ID}.

With the development of the sensing platform, state-of-the-art techniques, such as Lidar, high-resolution cameras, circular scanning millimeter-wave Radar, and laser range finder (LRF), have been applied on the BS and UAV to enable active sensing ability from VD. As a result, exciting works toward VD sensing-assisted link establishment are well underway, and such examples can be found in \cite{UAV_Learning_Beamforming1} and \cite{UAV_Learning_Beamforming2}. In addition, several pioneering works have preliminarily discussed the integration of sensing and communications (ISAC) for UAV networks. By exploiting the receivers’ motion parameters from the reflected echoes, lower overhead, and more accurate beam alignment have been realized \cite{UAV_ISAC1} \cite{UAV_ISAC3}. Nevertheless, a critical issue raised in the multi-UAV scenario is that the GUTI is not contained in the echoes or video frames \cite{cui2023specific} \cite{Com_Served_By_Sen}. Although the purpose-related receiver is usually determined by its physical characteristics rather than the GUTI \cite{Dual_Identity}, the GUTI is an indispensable ``ID card'' for verifying the physical link connection. The VD-only transmitter is more like a ``dumb man'' with complete visual ability since it does not know the GUTI of the intended receivers. Therefore, the GUTI and the physical characteristics of the intended receiver should be associated and maintained in the long-term flight to realize the following convenience. i) Given the physical characteristics of the intended receivers, it provides their corresponding GUTI, which ensures the fast establishment of links. ii) Given the GUTI of the intended receivers, it provides their corresponding physical characteristics, which ensures accurate beam alignment.

As per the above requirements of fast and accurate communication, the mmWave UAV network is supposed to be endowed with an advanced ability: \textit{\textbf{communicating with opening eyes}}. There naturally arises the need to associate the information obtained from both AD and VD. Nevertheless, it is not trivial to match what they hear and see since sometimes the physical characteristics of multiple UAVs tend not to differ much, inducing difficulty in distinguishing. The above issues and challenges motivate us to design a solution that could swiftly and accurately establish the links between the BS and the intended UAVs. 

In this paper, we offer a novel sensing-assisted beam management solution, which removes the tedious feedback and the resulting overhead for the multi-UAV cellular networks. Our scheme is relying on wireless and camera sensing. To be specific, we employ both the ISAC and computer vision techniques in downlink communication, where the video frames and the echo signals reflected by the UAVs are utilized to improve the performance of IA and beam tracking, respectively. As a result, the time consumption of the IA and beam tracking process is also significantly reduced. Besides, no downlink pilots and uplink feedback are required, and a significant matched-filtering gain is obtained in the SNR. The estimation errors are significantly reduced since most of the angular information is preserved, resulting in accurate beamforming. To provide accurate angle information for the downlink beamforming, we also developed the extended Kalman filtering (EKF) for tracking and prediction under the kinematic model of each UAV. The Mahalanobis distance (MD) is utilized to distinguish multiple UAVs in dynamic environments to calculate the similarity of two single physical characteristics. Aiming at solving the \textbf{Similarity} issue, the MDs are assigned dynamic weights according to the prevalence of characteristics. We also resort to the Jonker Volgenant algorithm for linear assignment problem (LAPJV), to associate the measurement and prediction by minimizing the similarity difference, which correctly aligns the specific beam toward the intended UAV. It avoids frequent maintenance of digital identities of UAVs during IM, and also solves the ``which is which'' issue. Consequently, the specific beam is accurately aligned toward the intended UAV, and the Kalman equation is updated. For clarity, we summarize the main contributions as follows.

\begin{itemize} 
\item We propose a novel sensing-assisted beam management solution for the multi-UAV cellular networks, which is the first solution that fully utilizes the ``VD'' information to improve the communication performance of UAV networks.
\item We propose an EKF method, and derive the linearization equation between observation and state transition models to realize accurate tracking.
\item We propose a novel dual identity association (DIA) solution aiming at distinguishing multiple UAVs in dynamic environments, which enables the specific beams to be accurately aligned toward the intended UAVs.
\end{itemize}

\section{system model} \label{System_model}
\subsection{motivation scenario}
\begin{figure}
	\centering
	\includegraphics[width=0.9\linewidth]{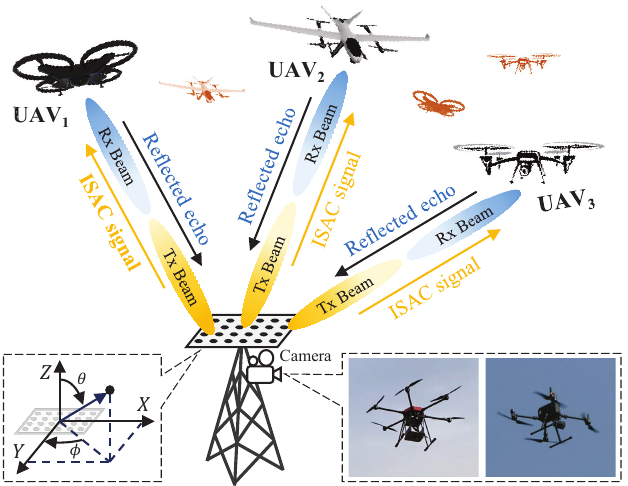}
	\caption{The scenario of the mmWave multi-UAV networks. The red and black UAVs are in IM and CM, respectively.}
	\label{Scenario}
\end{figure}

As shown in \textbf{Fig. \ref{Scenario}}, we consider a cellular-connected UAV network that the ground BS serves $K$ UAVs at mmWave frequencies. The BS is equipped with a mmWave massive MIMO (mMIMO) uniform planar array (UPA) consisting of $N_t$ transmit antennas and $N_{r_b}$ receive antennas. Moreover, the UPA is also assumed to be deployed at the bottom of UAVs. The DRX mechanism, which consists of IM and CM, is configured on BS and UAVs to save power. The CM and IM will be conducted alternately. 

The BS has the capability of sensing through computer vision and ISAC technology. By exploiting the independent transmit and receive antennas \cite{independent_transmit_and_receive}, the echoes for sensing can be received while maintaining uninterrupted downlink communications concurrently. The observation of UAV's physical characteristics is provided by ISAC in CM since the radio transmission during CM is the default capability. In IM, computer vision technology is adopted to achieve the same function since there is no data to transmit.

\begin{remark}
We assume that the self-interference issue at BS can be addressed by the separate transmit and receive UPA \cite{Self}, which may induce slight differences between the directions of transmitted and received beams when serving a near UAV. This issue can be tackled by compensating angles according to $D_1$, namely, the distance between the target and BS's receive UPA, and $D_2$, namely the distances between the separated receive and transmit arrays. If $D_1\gg D_2$, the difference in directions can be omitted since $D_1$ is approximately equal to the distance between the target and BS's transmit UPA. We will show the experiment results under perfect angle compensations, and designate a more detailed analysis of it as our future work.
\end{remark}

\subsection{channel model}
The channel model in \cite{Channel_Los_Rician} is employed here, and the line of sight (LoS) probability is given by a generalized logistic function ${\rm P}_{\rm LoS}(\theta_{k,n})=A_3 +\frac{A_4}{1+e^{-A_1-A_2(90 - \theta_{k,n})}}$, where $A_1<0$, $A_2>0$, $A_4>0$, and $A_3= 1-A_3$ are constants specified by the practical environment. $\theta_{k,n} = \frac{180}{\pi}{\rm arctan}(d_{k,n}^{\rm V}/d_{k,n}^{\rm H})$ is the elevation angle from the UAV to the BS shown in Fig. \ref{Scenario}, where $d_{k,n}^{\rm H}$ and $d_{k,n}^{\rm V}$ are the horizontal distance and vertical distance between BS and the $k$th UAV at the $n$th slot. With the locations of the BS and the $k$th UAV denoted as $\mathbf{p}_b$ and $\mathbf{p}_{k,n}=[{\rm p}_{k,n}(1),{\rm p}_{k,n}(2),{\rm p}_{k,n}(3)]^T$, we have their 3D distance $d_{k,n} = ||\mathbf{p}_{k,n}-\mathbf{p}_b||$.

If the LoS path exists, the communication channel can be modeled as 
$\textbf{h}_{k,n}=
\sqrt{{\rm K_R}}\textbf{h}_{k,n}^{\rm L}/\sqrt{{\rm K_R}+1}
+
\textbf{h}_{k,n}^{\rm N}
/\sqrt{{\rm K_R}+1}$,
where
$\textbf{h}_{k,n}^{\rm L}=\beta
e^{j2\pi\gamma t}
\delta(t-\tau)
\textbf{u}
\textbf{a}^H$
and
$\textbf{h}_{k,n}^{\rm N} = \sum\nolimits_{c=1}^{C}
\sum\nolimits_{p=1}^{P}
\tilde{\beta}_{c,p}
e^{j2\pi\tilde{\gamma}_{c,p}t}
\delta(t-\tilde{\tau}_{c,p})
\tilde{\textbf{u}}_{c,p}
\tilde{\textbf{a}}_{c,p}^H$
are the LoS and NLoS components of $\textbf{h}_{k,n}$, respectively. 
Symbols ${\rm K_R}$, $C$ and $P$ denote the Ricean K-factor, the number of clusters, and the number of paths in each cluster, respectively. We remark here that ${\rm q}$ and $\tilde{{\rm q}}_{c,p}$ respectively represent the parameter ``${\rm q}$'' of the LoS path and the $p$-th NLoS path in the $c$-th cluster. 
Symbols $\gamma$ and $\tau$ are the Doppler shift and delay of paths, and we have $\gamma_{k,n} = \mathbf{v}_{k,n}^T(\mathbf{p}_{k,n}-\mathbf{p}_b) f_c/(c||\mathbf{p}_{k,n}-\mathbf{p}_b||)$, where $\mathbf{v}_{k,n}=[{\rm v}_{k,n}(1),{\rm v}_{k,n}(2),{\rm v}_{k,n}(3)]^T$ denotes the velocity of the $k$th UAV, and $f_c$ and $c$ are the carrier frequency and the speed of light, respectively. $\beta$ and $\tilde{\beta}_{c,p}$ are the zero-mean Gaussian fading coefficients with variance $\sigma_{\beta}$ and $\sigma_{\tilde{\beta}_{c,p}}$.

Moreover, according to \cite{Los_P}, when the UAV communicates with the BS at different altitudes, different Rician factors should be adopted to efficiently characterize the channel, and ${\rm K_R}$ is modeled by an exponential function of $\theta_{k,n}$, i.e., ${\rm K_R}(\theta_{k,n})=B_1e^{B_2(\pi/2-\theta_{k,n}^R)}$, where $\theta_{k,n}^R$ is a radian version of $\theta_{k,n}$, $B_1$ and $B_2$ are constant coefficients determined by the specific environment. Then we have ${\rm K_R}_{\rm min}\le{\rm K_R}(\theta_{k,n})\le{\rm K_R}_{\rm max}$, where ${\rm K_R}_{\rm min}=B_1$ and ${\rm K_R}_{\rm max}=B_1e^{B_2*\pi/2}$.

The array steering vectors of the BS's transmit UPA is given by
\begin{equation}\label{Steering_Vector}
	\begin{aligned}
		&[\mathbf{a}]_{(n_{\rm y}-1)N_{\rm x}+n_{\rm x}}=\frac{1}{\sqrt{N_t}}e^{j\pi {\rm sin}\theta[(n_{\rm x}-1){\rm cos}\phi+(n_{\rm y}-1){\rm sin}\phi]},
	\end{aligned}
\end{equation}
where we assume the UPA has half-wavelength antenna spacing. 
Symbols $n_{\rm x}=1,...,N_{\rm x}$, $n_{\rm y}=1,...,N_{\rm y}$ denote the number of antenna along axis $x$ and $y$. Symbols $\phi,\theta$ in $\textbf{a}(\phi,\theta)$ denote the azimuth direction-of-departure (ADoD) and the elevation DoD (EDoD). The steering vector of the UAV’s receive UPA, namely $\textbf{u}(\phi',\theta')$, is similarly defined as $\textbf{a}(\phi,\theta)$ with $N_{k,r_u}$ receive antennas on the UAV, where $\phi',\theta'$ denote the azimuth direction-of-arrival (ADoA) and the elevation DoA (EDoA).

\subsection{signal and communication model}

By denoting the multi-beam ISAC signals toward $K$ UAVs as $\textbf{s}_n(t)=[s_{1,n}(t),…,s_{K,n}(t)]^T\in\mathbb{C}^{K\times1}$, the transmitted
signals are given by
\begin{equation}\label{Signal}
		\tilde{\textbf{s}}_n(t)=\mathbf{F}_n\textbf{s}_n(t)=\sum\nolimits_{k=1}^{K}\mathbf{f}_{k,n}s_{k,n}(t)\in\mathbb{C}^{N_t\times1},
\end{equation}
where $N_t=N_{tx}\times N_{ty}$ denotes the number of transmit antennas of the BS. The $k$th column of the beamforming matrix $\mathbf{F}_n$ is denoted as $\mathbf{f}_{k,n}=\mathbf{a}(\hat{\phi}_{k,n},\hat{\theta}_{k,n})$, where $\hat{\phi}$ and $\hat{\theta}$ are the estimated angles. 

The $k$th UAV receives the downlink signal from the BS via a receive beamformer $\mathbf{w}_{k,n}$, namely
\begin{equation}\label{Received_Signal_Full}
			r_{k,n}(t)=\kappa\sqrt{p_{k,n}}\mathbf{w}_{k,n}^H\textbf{h}_{k,n}\sum\nolimits_{k=1}^{K}\mathbf{f}_{k,n}s_{k,n}(t)+\mathbf{z}_r(t),
\end{equation}
where $\kappa = \sqrt{N_tN_{k,r_u}}$. The number of receive antennas of the $k$th UAV is denoted as $N_{k,r_u}$. For notational simplicity, we assume $N_{1,r_u} = ... = N_{K,r_u} = N_{r_u}$. $p_{k,n}$ represents the transmit power, and $\mathbf{z}_r(t)$ is the zero-mean Gaussian noise (hereinafter referred to as noise) with variance $\sigma_r^2$. 

Recall that the mMIMO UPA is adopted at the BS, the formulated beams will be adequately narrow since the steering vectors with different angles are nearly orthogonal. It can be mathematically expressed by \cite{mMIMO_Theory}
\begin{equation}\label{Lemma}
	|\mathbf{a}^H(\phi,\theta)\mathbf{a}(\phi^{\prime},\theta^{\prime})|\to0,\forall \theta\neq\theta^{\prime},\phi\neq\phi^{\prime},N_t\to\infty.
\end{equation}
Given the fact that there are safety distances between UAVs, namely $\theta_{k,n}\neq\theta_{k^{\prime},n}$, $\phi_{k^{\prime},n}\neq\phi_{k^{\prime},n}$, $\forall k\neq k^{\prime}$, we thus have $\mathbf{a}_{k,n}^H\mathbf{f}_{k',n} \approx0$, $\forall k\neq k^{\prime}$. Consequently, it yields $\mathbf{a}_{k,n}^H\sum\nolimits_{\tilde{k}=1}^{K}\mathbf{f}_{\tilde{k},n} \approx\mathbf{a}_{k,n}^H\mathbf{f}_{k,n}$, which means that the inter-beam interference between different UAVs in the downlink beams could be negligible. The signal received by the $k$th UAV can thus be approximated as
\begin{equation}\label{Received_Signal}
		r_{k,n}(t)=\kappa\sqrt{p_{k,n}}\mathbf{w}_{k,n}^H\textbf{h}_{k,n}\mathbf{f}_{k,n}s_{k,n}(t)+\mathbf{z}_r(t),
\end{equation}

Assuming that the ISAC signal has a unit power, then the receive SNR is expressed as
\begin{equation}\label{Received_SNR}
\Gamma_{k,n}=p_{k,n}|\kappa\alpha\mathbf{w}_{k,n}^H\mathbf{b}_{k,n}\mathbf{a}_{k,n}^H\mathbf{f}_{k,n}|^2/\sigma_r^2.
\end{equation}
Since the UAV knows its own velocity $\mathbf{v}_{k,n}$ and location $\mathbf{p}_{k,n}$, the Doppler shift $\gamma_{k,n}$ can be readily compensated at the UAV’s receiver. The average achievable rate of all the $K$ UAVs is given as $R_n=\frac{1}{K}\sum_{k=1}^{K}{{\rm log}_2\left(1+\Gamma_{k,n}\right)}$.

\subsection{measurement model}
\subsubsection{Radar-based sensing}
The echoes reflected by all UAVs are received by the BS, which can be formulated as
\begin{equation}\label{Echo_Total}
	\begin{aligned}
\mathbf{c}_n(t)=\tilde{\kappa}\sum\nolimits_{k=1}^{K}&\sqrt{p_{k,n}}\beta_{k,n}e^{j2\pi\mu_{k,n}t}\mathbf{u}'_{k,n}\\&\times\mathbf{a}_{k,n}^H\tilde{\textbf{s}}_{n}(t-\tau_{k,n})+\mathbf{z}_c(t).
  	\end{aligned}
\end{equation}
Here, symbols $\tilde{\kappa}=\sqrt{N_tN_{r_b}}$, $N_{r_b}$, $\beta_{k,n}$, $\mu_{k,n}$, and $\tau_{k,n}$ denote the array gain factor, the number of receive antennas, the reflection coefficient, the Doppler frequency, and the time-delay, respectively. The steering vector $\mathbf{u}'(\phi_{k,n}, \theta_{k,n})$ of BS's receive UPA is similarly defined as Eq. \eqref{Steering_Vector} with $N_{r_b}$ antennas. 

Note that Eq. \eqref{Lemma} also reveals that $|\mathbf{b}_{k,n}^H\mathbf{u}'_{k',n}| \approx0$, $\forall k\neq k^{\prime}$, namely the inter-beam interference between different UAVs in the uplink echoes can be omitted. Here, $\mathbf{b}_{k,n}(\hat{\phi},\hat{\theta})$, which is similarly defined as Eq. \eqref{Steering_Vector} with $N_{r_b}$ antennas, can be determined by predicting angles based on the estimations from the $n-1$th slot. The BS can distinguish different UAVs in terms of angle-of-arrivals for independent processing. Consequently, the received echo at the BS from the $k$th UAV, denoted by $\mathbf{c}_{k,n}(t)$, can be extracted from Eq. (\ref{Echo_Total}) via a spatial filtering process, i.e., multiplying an item of $\mathbf{b}_{k,n}^H$ with $\mathbf{c}_n(t)$, which can thus be approximated as
\begin{equation}\label{Echo1}
	\begin{aligned}
	&\mathbf{c}_{k,n}(t)=\mathbf{b}_{k,n}^H\mathbf{c}_n(t)\\=&\tilde{\kappa}\sqrt{p_{k,n}}\beta_{k,n}e^{j2\pi\mu_{k,n}t}\mathbf{a}_{k,n}^H\mathbf{f}_{k,n} s_{k,n}(t-\tau_{k,n})+\mathbf{z}_{k,n}(t),
 	\end{aligned}
\end{equation}
where $\mathbf{z}_{k,n}(t)$ represents the noise with variance $\sigma^2$.

The $\mu_{k,n}$ and $\tau_{k,n}$ can be estimated by matched-filtering $\mathbf{c}_{k,n}(t)$ with a Doppler-shifted delayed version of $s_{k,n}(t)$ \cite{Variance}. Therefore, $\mathbf{p}_{k,n}$ and $\mathbf{v}_{k,n}$ are measured by
\begin{equation}\label{Range_and_Velocity}
	\left\{
	\begin{aligned}
		&\hat{\tau}_{k,n}=\frac{2||\mathbf{p}_{k,n}-\mathbf{p}_b||}{c}+z_\tau \\
		&\hat{\mu}_{k,n}=\frac{2\mathbf{v}_{k,n}^T(\mathbf{p}_{k,n}-\mathbf{p}_b) f_c}{c||\mathbf{p}_{k,n}-\mathbf{p}_b||}+z_f
	\end{aligned}
	\right.,
\end{equation}
where $z_\tau$ and $z_f$ denote the noise with variance $\sigma_1^2$ and $\sigma_2^2$, respectively. Note that the estimation of the reflection coefficient can be realized via ${\hat{\beta}}_{k,n}=\xi/\hat{\tau}_{k,n}c$, which depends on the radar cross-section $\xi$ of UAV. By compensating $\mathbf{c}_{k,n}(t)$ with the estimation and normalizing the results by $p_{k,n}$ and the matched-filtering gain $G$, we obtain a compact measurement model for the angles $\theta$ and $\phi$ as
\begin{equation}\label{Echo2} 
	{\tilde{c}}_{k,n}\!=\tilde{\kappa}\beta_{k,n}\mathbf{a}_{k,n}^H\mathbf{f}_{k,n}+\mathbf{\tilde{z}}_{k,n}.
\end{equation}

The variance of the measuring noise $\mathbf{\tilde{z}}_{k,n}$ is denoted as $\sigma_3^2$. According to \cite{Variance}, the variance $\sigma_i^2, i=1\sim3$ are inversely proportional to the receive SNR of $\mathbf{c}_{k,n}(t)$, and we thus assume that $\sigma_i^2=a_i^2\sigma^2/(GN_tN_{r_b}|\beta_{k,n}|^2|\mathbf{a}_{k,n}^H(\phi,\theta)\mathbf{f}_{k,n}|^2p_{k,n}),i=1,2$, and $\sigma_i^2=a_i^2\sigma^2/(Gp_{k,n}),i=3$, where $a_i, \forall i$ depend on the system configuration, etc.

\subsubsection{Vision-based sensing}\label{VISION-BASED SENSING}
\begin{figure}
	\centering
	\includegraphics[width=0.7\linewidth]{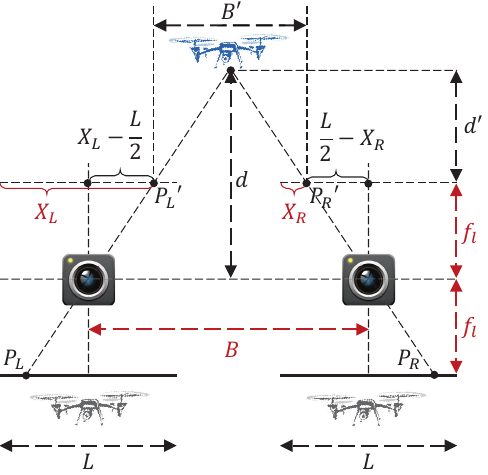}
	\caption{The schematic diagram of binocular stereo vision parallel top-down geometric imaging.}
	\label{Vision}
\end{figure}

In addition to measuring the velocity and distance measurement based on echo signal, we also exploit computer vision to activate the sensing capability in IM. Specifically, you only look once (YOLO) is developed to detect the presence of receivers, and the binocular camera is utilized to measure the relative distance. Two cameras are primarily used to obtain different information about the same target from two perspectives, thereby obtaining the depth information from the image's depth of field of view. \textbf{Fig. \ref{Vision}} is a schematic diagram of binocular stereo vision parallel top-down geometric imaging. From a similar relationship, we have $B^{\prime}/d^{\prime}=B/d$. It can be rewritten as $[B-(X_{L}-\frac{L}{2})-(\frac{L}{2}-X_{R})]/(d-f_l)=B/d$ and therefore we get $d= Bf_l/(X_{L}-X_{R})$, where $B$ is the optical center distance of the two cameras and $f_l$ is the camera focal length. The disparity $X_L-X_R$ describes the position difference of a spatial point mapped to the projection point on the left and right image planes of the binocular camera. By recording the distance estimated against time in a list, the relative velocity of the detected UAVs can be calculated.

There are numerous solutions to solve the issue when the LoS paths and images are blocked by occlusion. For instance, computer vision techniques can handle partial occlusion, and the sensing performance under severe partial occlusion can be remarkably improved by a thermal imaging camera. Besides, one of the promising solutions for complete occlusion is the ``around-the-corner radar'', and the terahertz radars can also tackle this issue if the distance is close enough. In addition, there are also numerous factors that affect the accuracy of camera sensing, e.g., the attitude of the target, sensing algorithms, number of cameras, and bias of camera orientation.

Given the page limitations and the extensive research on occlusion and camera sensing that already exists, we choose not to discuss them to avoid alleviating the focus of sensing-assisted communication to camera sensing or tracking under occlusion, and designate a more detailed discussion as future work. We will show the numerical results of the tracking performance under the error results we obtained in real-world experiments, which are based on YOLO version 5 and DJI's Zenmuse H20 camera.

\section{Framework}
\begin{figure*}
	\centering
	\includegraphics[width=0.75\linewidth]{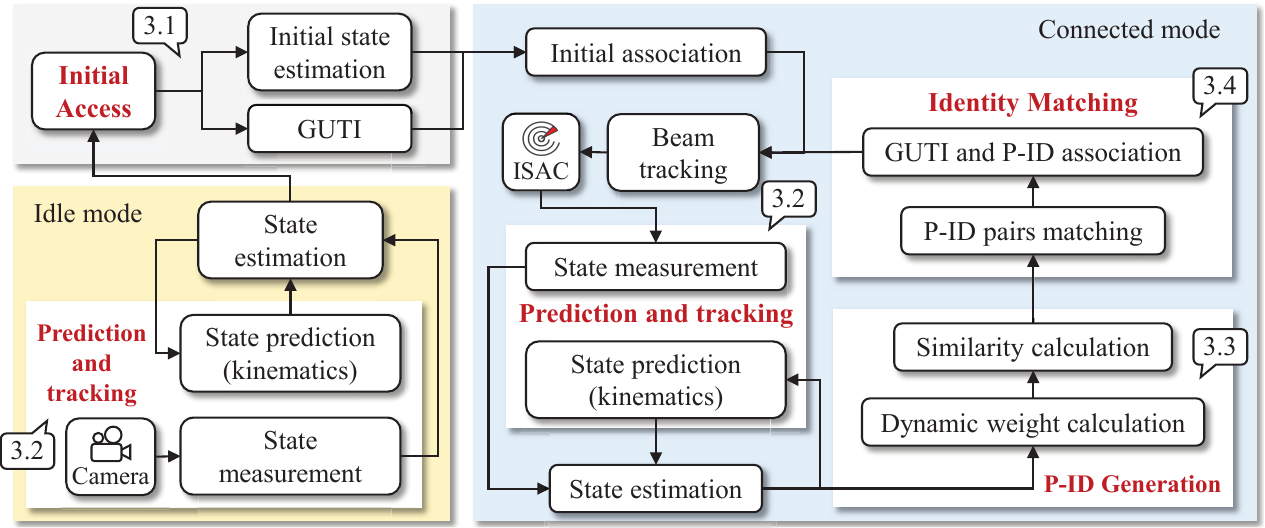}
	\caption{The framework of the proposed sensing-assisted beam management solution.}
	\label{Framework}
\end{figure*}
As shown in \textbf{Fig. \ref{Framework}}, the framework of our proposed sensing-assisted beam management solution is mainly composed of four modules: i) initial access, ii) prediction and tracking, iii) physical identity (P-ID) generation, and iv) identity matching.

\subsection{initial access}
Recall that the DRX mode is frequently utilized due to the limited energy of the UAV network, which is switched periodically between IM and CM. When the network is switching from IM to CM, the IA process is accelerated by utilizing the vision-based sensing technique described in Sec. \ref{VISION-BASED SENSING}. The position of UAVs and BS is observed by their continuous visual tracking results. As a result, the optimal beam direction of BS's transmit UPA is not iteratively determined via conventional beam training. On the contrary, it will be swiftly selected from a potential beam set $\tilde{\mathcal{S}}_{\mathrm{B}}$, which is determined by the UAV's velocity and distance derived from the YOLO technique. This potential beam set is selected from the complete beam set of BS, which is defined as
\begin{equation}
	\begin{aligned} 
		&\mathcal {S}_{\mathrm{B}}=\left\{ \mathbf {a}(\bar{\phi }(m),\bar{\theta }(m^{\prime }))| \bar{\phi }(m)=\frac{m-1}{2^{\frac{Q_{\mathrm{B}}}{2}}}\pi, \right.\\ &\left.\bar{\theta }(m^{\prime})=\frac{m^{\prime }-1}{2^{\frac{Q_{\mathrm{B}}}{2}}}\pi,m,m^{\prime }=1,2,\dots,2^{\frac{Q_{\mathrm{B}}}{2}}\right\},
	\end{aligned}
\end{equation}
where $Q_{\mathrm{B}}$ is the number of bits that control the phase of the BS's UPA. Besides, according to the location relative to the BS, UAVs also select a set of potential beam directions $\tilde{\mathcal{S}}_{\mathrm{U}}$ from the complete beam set
\begin{equation}
	\begin{aligned} 
		&\mathcal {S}_{\mathrm{U}}=\left\{ \mathbf {b}(\check{\phi }(m),\check{\theta }(m^{\prime }))| \check{\phi }(m)=\frac{m-1}{2^{\frac{Q_{\mathrm{U}}}{2}}}\pi, \right.\\ &\left.\check{\theta }(m^{\prime })=\frac{m^{\prime }-1}{2^{\frac{Q_{\mathrm{U}}}{2}}}\pi,m,m^{\prime }=1,2,\dots,2^{\frac{Q_{\mathrm{U}}}{2}}\right\},
	\end{aligned}
\end{equation}
where $Q_{\mathrm{U}}$ is the number of bits that control the phase of the UAV's UPA. 

The potential beam set $\tilde{\mathcal{S}}_{\mathrm{B}}$ and $\tilde{\mathcal{S}}_{\mathrm{U}}$ are a subset of the complete set $\mathcal{S}_{\mathrm{B}}$ and $\mathcal{S}_{\mathrm{U}}$, respectively. With the vision-based sensing result in hand, i.e., the velocity and distance of a UAV, the azimuth and elevation angles can be obtained. These angles will further be utilized to find an angle pair $\bar{\phi}(m)$ and $\bar{\theta}(m^{\prime})$ in $\mathcal{S}_{\mathrm{B}}$, which is the closest angles to them. Based on this angle pair, $\lceil (|\tilde{\mathcal{S}}_{\mathrm{B}}|-1)/2\rceil$ and $|\tilde{\mathcal{S}}_{\mathrm{B}}|-\lceil (|\tilde{\mathcal{S}}_{\mathrm{B}}|-1)/2\rceil-1$ angles are selected in ascending and descending order in $\mathcal{S}_{\mathrm{B}}$ respectively, and the potential beam set $\tilde{\mathcal{S}}_{\mathrm{B}}$ will be formed in this way. At the UAV side, the potential beam set $\tilde{\mathcal{S}}_{\mathrm{U}}$ can also be formed in the same way.

The proposed IA procedure is explained as follows. 1) PSS detection: The primary synchronization signal (PSS) is sent by the BS once every $T_p$ ms, and the BS uses the potential beams in the set $\tilde{\mathcal{S}}_{\mathrm{B}}$. Meanwhile, the UAV uses beams in the beam set $\tilde{\mathcal{S}}_{\mathrm{U}}$ to receive the PSS. 2) Random access (RA) preamble transmission: The UAV transmits an RA preamble once every $T_p$ ms, with the beam that corresponds to the highest SNR in the PSS detection stage. The BS sequentially uses beams in the set $\tilde{\mathcal{S}}_{\mathrm{B}}$ to receive the RA preamble. If the highest SNR surpasses the threshold, the corresponding beam is selected for transmission. 3) Connection: If the RA preamble is successfully detected, further connection requests and channel scheduling will be conducted. Here we restrict the cardinality as $|\tilde{\mathcal{S}}_{\mathrm{B}}|=S_1$ and $|\tilde{\mathcal{S}}_{\mathrm{U}}|=S_2$, and they also satisfy $S_1\ll2^{Q_{\mathrm{B}}}$ and $S_2\ll2^{Q_{\mathrm{U}}}$. So, the delay of the proposed IA procedure can be analytically characterized as $|\tilde{\mathcal{S}}_{\mathrm{B}}|\times(|\tilde{\mathcal{S}}_{\mathrm{U}}|+1)\times T_p$. The only parameters that affect IA delay are the cardinality of the sets $\tilde{\mathcal{S}}_{\mathrm{B}}$ and $\tilde{\mathcal{S}}_{\mathrm{B}}$. Taking $\tilde{\mathcal{S}}_{\mathrm{B}}$ as an example, it is a set extracted from a complete set $\mathcal{S}_{\mathrm{B}}$, and the cardinality of $\tilde{\mathcal{S}}_{\mathrm{B}}$ depends on two aspects: 1) the confidence level of the camera-sensing result when capturing the UAV, and 2) the number of antennas of the BS's UPA.

After IA, the UAVs' GUTI and motion parameters, i.e., $\phi_{k,0}$, $\theta_{k,0}$, $d_{k,0}$ and $v_{k,0}$, are successfully obtained by the BS and the BS's location information is also successfully obtained by the UAV. As a result, the beam will be initially aligned and the association between GUTI and initial physical characteristics will be obtained since they are fed back from specific UAVs. This initial association is the precondition for the subsequent identity matching since the conventional feedback is replaced by echo signals or video frames, from which we can only obtain physical characteristics rather than GUTI.

\subsection{State Prediction and Tracking}
To enable fast communication, beams need to be aligned at fleeting opportunities in high-mobility UAV networks. Different from the communication-only feedback-based beam alignment methods, the mutual prediction and tracking of the BS and the UAVs are conducted in our proposal. It will continue for a long period of time $T$, which is discretized into several small time slots $\Delta T$.

In CM, the ISAC signal transmitted by the BS is received by the UAV's UPA and also reflected by the fuselage. On the BS side, the echos are exploited to measure the UAVs' motion parameters at the $n-1$th epoch by Eq. \eqref{Range_and_Velocity} and \eqref{Echo2}. On the UAV side, the receive direction is calculated according to their own real-time mobility and the location information of BS, which has been acquired in the IA process.

In IM, the ISAC functionality is absent since there is no requirement for wireless transmission. The video frames captured at both the BS side and the UAV side, as the substitute for radar sensing, will provide the measurement of motion parameters at the $n-1$th epoch. 

The echoes and video frames are utilized to refine the predicted parameters at the $n-1$th epoch to obtain the estimations $\phi_{k,n-1}$, $\theta_{k,n-1}$, $d_{k,n-1}$ and $v_{k,n-1}$. To avoid the measurement becoming outdated, the refined state parameters are then used as the inputs of the predictor for the $n$th epoch. Both the BS and UAVs will perform one-step prediction by using the kinematic equations. A detailed derivation of the kinematic state evolution model will be presented in Sec. \ref{EKF}.

The BS (and UAVs) will formulate transmit (receive) beamformer $\mathbf{f}_{k,n}$ ($\mathbf{w}_{k,n}$) based on $\mathbf{\hat{\theta}}_{k,n|n-1}$ and $\mathbf{\hat{\phi}}_{k,n|n-1}$ by using the predictions at the $n$th epoch. As a result, the beams between BS and UAVs will be aligned once obtaining accurate predictions.
	
\subsection{P-ID Generation}
Due to the \textbf{Similarity} issue, the motion parameters of multiple UAVs may be similar sometime or somewhere. Therefore, after obtaining the measurement and prediction of the UAVs' states, the BS further generates the distinguishable P-ID to make a unique physical identification for each UAV. Specifically, it denotes $\textbf{PF}_n=\left\{\textbf{\textbf{pf}}_{k,n}, k=1,...,K\right\}$ as the observed physical characteristic vector of $K$ UAVs, where $\textbf{\textbf{pf}}_{k,n}(m)$ denotes the $m$th physical characteristic ($m=1,...,M$) of the $k$th UAV. The P-IDs are established by collecting all the observable and distinguishable physical characteristics and formulating dynamic weights based on their prevalence. More details will be presented in Sec. \ref{RE_Weight_Similarity}.
\begin{figure}
	\centering
	\includegraphics[width=0.95\linewidth]{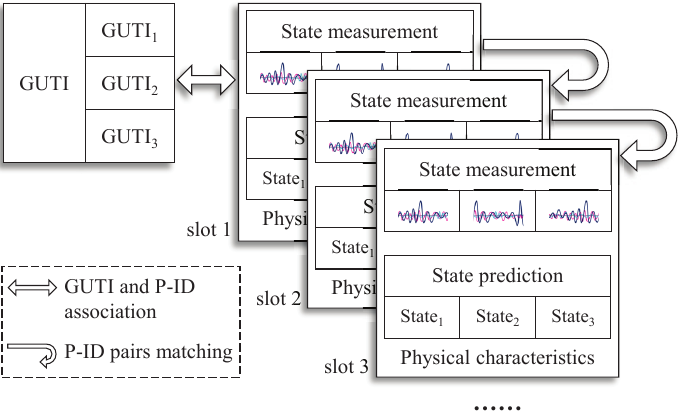}
	\caption{The illustration of P-ID pairs matching, and GUTI and P-ID association.}
	\label{Matching_concept}
\end{figure}
\begin{remark}
It should be noted that any distinguishable physical characteristic is included in the connotation of P-ID. For instance, the unique micro-Doppler frequency is caused by blade numbers and rotor speeds.
\end{remark}

\subsection{Identity Matching}
Recall that the data sent to specific UAVs should be carried on the designated beams. Nevertheless, the UAVs' GUTI cannot be obtained from radar echoes. Therefore, in our uplink feedback-free scheme, the GUTI and motion parameters of UAVs are kept matched continuously on the BS side to enable accurate communication. As shown in \textbf{Fig. \ref{Matching_concept}}, the initial association result of GUTI and P-ID has been obtained in the IA stage, and the P-ID information continuously comes from radar-based sensing. Regardless of where the P-ID pairs of any two adjacent time slots come from, the accurate matching of GUTI and P-ID can be realized in the long term once P-ID pairs are iteratively associated. The details scheme for associating the P-IDs obtained at different epochs will be discussed in Sec. \ref{Matching_ID}. Once the P-ID and GUTI are successfully matched, a specific beam will be aligned toward the intended UAV, and the EKF will be correctly updated, which ensures the accuracy of beam alignment.
	
\section{Proposed EKF and DIA}
In what follows, we will detail the technique we designed for the proposed EKF and DIA approaches and the discussion about computational complexity.

\subsection{Extended Kalman Filtering}\label{EKF}

In this subsection, a Kalman filtering scheme is proposed for beam prediction and tracking. Due to the nonlinearity in the measurement functions, the linear Kalman filtering can not be utilized directly. We thus develop an EKF method that performs linearization for nonlinear measurement. 

Take the radar-based sensing as an example, the state variables and measured vectors are denoted as $\boldsymbol{x}_{k,n}=[\,\mathbf{p}_{k,n},\mathbf{v}_{k,n},\mathbf{\textit{a}}_{k,n}]^T$ and $\boldsymbol{y}_{k,n}=[\tau_{k,n},\mu_{k,n},\tilde{\mathbf{c}}_{k,n}]^T$. Assuming UAV moves with a constant acceleration mobility model, the models of state evolution and measurement can be given by
\begin{equation}\label{Motion_model}
	\left\{
	\begin{aligned}
		&{\rm State\ evolution\!:\ }\boldsymbol{x}_n=\textbf{G}\boldsymbol{x}_{n-1}+\boldsymbol{u}_{n-1}\\
		&{\rm Measurement\!:\ }\boldsymbol{y}_n=\textbf{H}(\boldsymbol{x}_{n})+\boldsymbol{z}_{n}
	\end{aligned}
	\right.,
\end{equation}
where $\textbf{G}=[\textbf{I}_{3\times3},\Delta T\cdot\textbf{I}_{3\times3},\frac{\Delta T}{2}\cdot\textbf{I}_{3\times3}; \textbf{0}_{3\times3},\textbf{I}_{3\times3},$$\Delta T\cdot\textbf{I}_{3\times3};\textbf{0}_{3\times3},\textbf{0}_{3\times3},\textbf{I}_{3\times3}]$, and $\textbf{H}(\cdot)$ is defined as Eq. \eqref{Range_and_Velocity} and \eqref{Echo2}. $\boldsymbol{u}$ and $\boldsymbol{z}$ are noises with covariance matrices as $\textbf{Q}_s={\rm diag}(\sigma_{p(i)}^2,\sigma_{v(i)}^2,\sigma_{a(i)}^2), i=1,2,3$ and $\textbf{Q}_m={\rm diag}(\sigma_{1}^2,\sigma_{2}^2,\sigma_{3}^2)$.

In order to linearize the measurement models, let's denote $\eta(\theta,\phi)=\sqrt{N_tN_{r_b}}\beta\mathbf{a}^H\!(\phi,\!\theta)\mathbf{a}(\hat{\phi},\!\hat{\theta})$. The Jacobian matrix for $\textbf{H}(\boldsymbol{x})$ is given as $[\dfrac{\partial \textbf{H}}{\partial \boldsymbol{x}},\textbf{0}_{3\times3}]$, where
\begin{equation}\label{jacobian}
	\dfrac{\partial \textbf{H}}{\partial \boldsymbol{x}}=\!\left[\begin{matrix}m(1)\!\!\!&m(2)\!\!\!&m(3)\!\!\!&0\!\!\!&0\!\!\!&0\\b(1)\!\!\!&b(2)\!\!\!&b(3)\!\!\!&\tilde{m}(1)\!\!\!&\tilde{m}(2)\!\!\!&\tilde{m}(3)\\q(1)\!\!\!&q(2)\!\!\!&q(3)\!\!\!&0\!\!\!&0\!\!\!&0\end{matrix}\right],
\end{equation}
where 
\begin{equation}
		\left\{
\begin{aligned}
&m(i)=\frac{2{\rm p}(i)}{c\sqrt{\mathbf{p}^T\mathbf{p}}},\ \tilde{m}(i)=f_cm(i)\\ 
&b(i)=\frac{2f_c[{\rm v}(i)\mathbf{p}^T\mathbf{p}-{\rm p}(i)\mathbf{p}^T\mathbf{v}]}{c(\mathbf{p}^T\mathbf{p})^\frac{3}{2}}\\ 
&q(i)=\frac{\partial\eta}{\partial\theta}\frac{{\rm p}(i){\rm p}(3)}{\mathbf{p}^T\mathbf{p}\sqrt{\mathbf{p}^T\mathbf{p}-{\rm p}^2(3)}}
\end{aligned}
\right., i=1,2,3.
\end{equation}
The partial derivative of $\boldsymbol{\eta}$ with respect to $\theta$ is given by
\begin{equation}\label{partial_eta}
	\frac{\partial\eta}{\partial\theta} = -\beta\sqrt{\frac{N_{r_b}}{N_{t}}}\sum\limits_{n_{tx} = 1}^{N_{tx}}\sum\limits_{n_{ty} = 1}^{N_{ty}}\frac{\chi(\hat{\theta},\hat{\phi})}{\chi(\theta,\phi)}\frac{\partial ln(\chi(\theta,\phi))}{\partial\theta},
\end{equation}
where $\chi(\theta,\phi)=e^{j\pi sina[(n_{ty}-1)cosb+(n_{tx}-1)sinb]}$.

We are now ready to present the EKF procedure, and the state prediction and tracking steps are summarized as follows.

1) State prediction: 

$\hat{\boldsymbol{x}}_{n|n-1}=\textbf{G}\boldsymbol{x}_{n-1}$.

2) Linearization: 

${\rm \textbf{H}}_n=\left.\frac{\partial \textbf{H}}{\partial \boldsymbol{x}}\right|_{\boldsymbol{x}=\boldsymbol{\hat{x}}_{n|n-1}}$.

3) Prediction of the mean squared error matrix: 

${\rm \textbf{M}}_{n|n-1}={\rm \textbf{G}}_{n-1}{\rm \textbf{M}}_{n-1}{\rm \textbf{G}}_{n-1}^H+{\rm \textbf{Q}}_s$.

4) Calculation of the Kalman gain: 

${\rm \textbf{K}}_n={\rm \textbf{M}}_{n|n-1}{\rm \textbf{H}}_{n}^H({\rm \textbf{Q}}_{m}+{\rm \textbf{H}}_{n}{\rm \textbf{M}}_{n|n-1}{\rm \textbf{H}}_{n}^H)^{-1}$.

5) State tracking: 

$\hat{\boldsymbol{x}}_{n}=\hat{\boldsymbol{x}}_{n|n-1}+{\rm \textbf{K}}_n(\boldsymbol{y}_{n}-{\rm \textbf{H}}(\hat{\boldsymbol{x}}_{n|n-1}))$.

6) Updating the mean squared error matrix: 

${\rm \textbf{M}}_{n}=({\rm \textbf{I}}-{\rm \textbf{K}}_{n}{\rm \textbf{H}}_{n}){\rm \textbf{M}}_{n|n-1}$. 

By performing prediction and tracking iteratively, the BS can simultaneously sense and communicate with $K$ UAVs according to the optimal angles. To provide the best angles for the beamformer, the predicted location $\mathbf{\hat{p}}_{k,n}$ could be utilized to calculate the predicted angles by the unbiased converted measurements method, and more details can be found in our recent work \cite{Dual_Identity}.

\begin{remark}
While the UAV may have nonlinear mobility, e.g., random-based, time-based, path-based, group-based, and topology-based mobility, the nonlinearity of the UAV's flight is mainly affected by the time-varying acceleration. The motion model established in Eq. (\ref{Motion_model}) represents the irregular turning situation, namely all the mobility models can be regarded as a combination of Eq. (\ref{Motion_model}). In this case, the tracking accuracy and the beam prediction accuracy depend on the linearization error. The tracking accuracy can be further improved by observing the target's acceleration based on some transform, e.g., fractional Fourier transform.
\end{remark}

\subsection{Prevalence-based Weight for Similarity} \label{RE_Weight_Similarity}\label{Prevalence}
This subsection proposed a dynamic P-ID generation method to provide the Sec. \ref{Matching_ID} with a reliable similarity metric.

Given the prediction and estimation of UAV's states, we want to calculate $S_n(i,j)$, namely the similarity of the measurements $\boldsymbol{y}_{n}$ and their estimations $\hat{\boldsymbol{y}}_{n|n-1}$. Nevertheless, this is not trivial due to the \textbf{Similarity} issue. For instance, locations are more distinctive in a low-density network, while velocity has low importance in distinguishing when UAVs are in formation. This motivates us to dynamically compute the weight based on their prevalence.
	
\subsubsection{Mahalanobis Distance Calculation}
Note that the MD fully utilizes the covariance in the feature vectors, and consequently has the capability to avoid bias in any variable dimension. The MD of two feature vectors $\textbf{\textit{f}}_a$ and $\textbf{\textit{f}}_b$ is defined as $C(\textbf{\textit{f}}_a,\textbf{\textit{f}}_b)=\sqrt{\left(\textbf{\textit{f}}_a-\textbf{\textit{f}}_b\right)^T\Sigma^{-1}\left(\textbf{\textit{f}}_a-\textbf{\textit{f}}_b\right)}$,
where $\Sigma^{-1}$ is the inverted covariance matrix.

\subsubsection{Dynamic Weight Assignment}
We denote $\textbf{\textbf{pf}}_n(m)=\{\textbf{\textbf{pf}}_{k,n}(m), k=1,...,K\}$ as the vector of all measurements of the $m$th physical characteristic on $K$ UAVs. The weight of the $m$th physical characteristic is assigned as $w_{n}(m)=\frac{1}{K}\sum\nolimits_{k=1}^{K}P_{k,n}(m)$,
where $P_{k,n}(m)$ is the distinguishability of $\textbf{\textbf{pf}}_{k,n}(m)$, namely the probability that $\textbf{\textbf{pf}}_{k,n}(m)$ is different from other physical characteristics in $\textbf{\textbf{pf}}_n(m)$. Here we define $P_{k,n}(m)$ as
\begin{equation}\label{Distinguishability}
	\begin{aligned}
		P_{k,n}(m)&=\sum\limits_{j\neq k}C(\textbf{\textbf{pf}}_{k,n}(m), \textbf{\textbf{pf}}_{j,n}(m))\times\\&\prod\limits_{q\neq j,q\neq k}(1-C(\textbf{\textbf{pf}}_{k,n}(m), \textbf{\textbf{pf}}_{q,n}(m))),
	\end{aligned}
\end{equation}
where the term $C(\textbf{\textbf{pf}}_{k,n}(m), \textbf{\textbf{pf}}_{j,n}(m))\prod\nolimits_{q\neq j,q\neq k}(1-C(\textbf{\textbf{pf}}_{k,n}(m), \textbf{\textbf{pf}}_{q,n}(m)))$ denotes the probability that $\textbf{\textbf{pf}}_{k,n}(m)$ is the same as $\textbf{\textbf{pf}}_{j,n}(m)$ but different from the other measurements.
	
Based on Eq. \eqref{Distinguishability}, the physical characteristic with a high distinguishability will play a more important role in identifying UAVs. Then the similarity of $\textbf{\textbf{pf}}_{i,n}$ and $\textbf{\textbf{pf}}_{j,n}$ is defined as the harmonic mean of individuals
\begin{equation}\label{Similarity} 
			S_n(i,j)=\left\{\sum\limits_{m=1}^{M}\frac{w_n^{\prime}(m)}{C\left(\textbf{\textbf{pf}}_{i,n}(m), \textbf{\textbf{pf}}_{j,n}(m)\right)}\right\}^{-1}, 
\end{equation}
where $w_n^{\prime}(m)$ denotes the normalized weight, and $M$ is the number of observable physical characteristics. 

The above methods dynamically weigh physical characteristics via their prevalence. Besides, if two UAVs have high dissimilarities in most physical characteristics, their similarity will be low despite the large weights of other physical characteristics. As a result, the similarity matrix will be more convenient for P-ID pairs association. 
		
\subsection{P-ID Pairs Matching}\label{Matching_ID}
This subsection details the necessity and method for identity matching. 

Note that the original association of UAV's GUTI and P-ID has been obtained at the IA stage. In the subsequent process, the P-ID information is obtained by processing echo signals instead of communication feedback. However, because the GUTI is not contained in the echos, the BS should have the capability to distinguish UAVs by associating the subsequently measured P-ID with the GUTI obtained at the IA stage to realize the following two aims: i) correctly aligning the specific beam toward the intended UAV and ii) correctly updating the state estimation introduced in Sec \uppercase\expandafter{\romannumeral4}-A. This implies that each UAV's P-ID information must be accurately correlated in any two adjacent time slots. To tackle this issue, we propose an efficient P-ID pair-matching approach.

\begin{figure}
\centering
\includegraphics[width=1\linewidth]{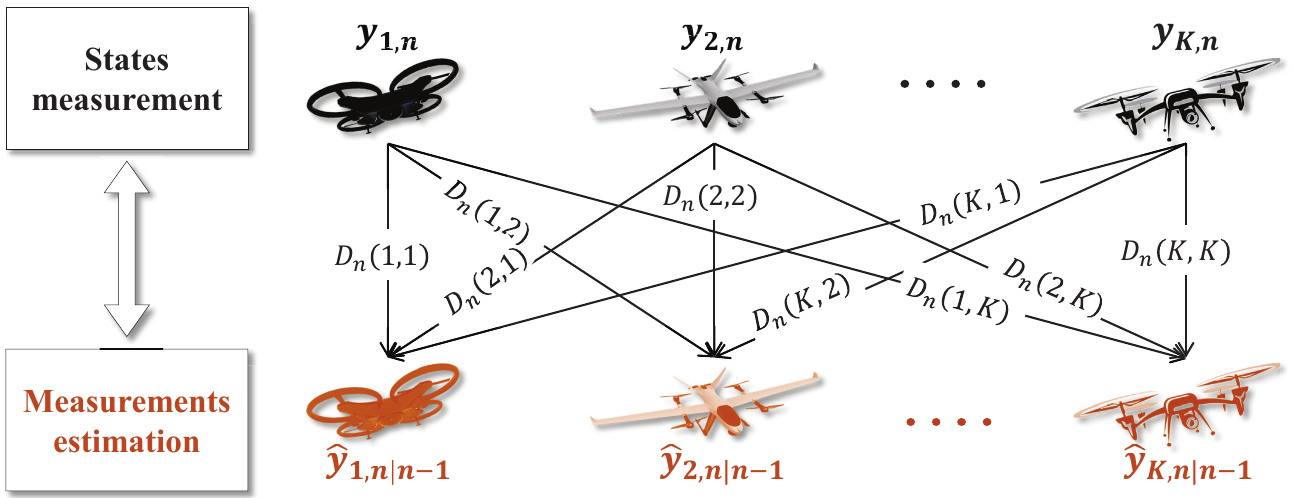}
\caption{Bipartite graph model for matching P-ID pairs.}
\label{Matching_Model}
\end{figure}

Recall that the BS detects $K$ UAVs and formulate measurements $\boldsymbol{y}_{k,n}, k=1,...,K$, we calculate all the measurement estimation of state predictions by $\hat{\boldsymbol{y}}_{j,n|n-1}=\textbf{H}(\hat{\boldsymbol{x}}_{j,n|n-1}),j=1,...,K$. The measurement $\boldsymbol{y}_{i,n}$ will not be ``far from" $\hat{\boldsymbol{y}}_{j,n|n-1}$, so we calculated their difference, namely the reciprocal of their similarity by $D_n(i,j)=S_n(i,j)^{-1}$, and then establish the bipartite graph matching model shown in \textbf{Fig. \ref{Matching_Model}}, where the edge's weight is defined as the matching cost $D_n(i,j),\forall i,j$.
	
\begin{figure*}
\centering
\includegraphics[width=1\linewidth]{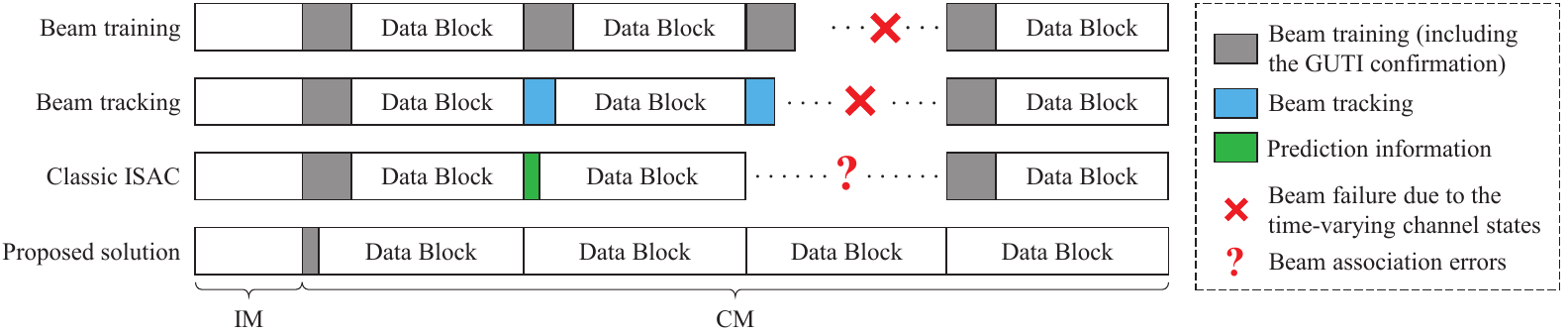}
\caption{The transmission block structure of the proposed solution.}
\label{Frame_Structure}
\end{figure*}

The optimization target of the P-ID pairs matching problem is to minimize the overall differences, i.e., $\min \sum\nolimits_{i=1}^{K}\sum\nolimits_{j=1}^{K}A_n(i,j)D_n(i,j)$,
which is constrained to $\sum\nolimits_{i=1}^{K}A_n(i,j)=1$ and $\sum\nolimits_{j=1}^{K}A_n(i,j)=1$, $i,j=1,...,K$, namely any $\boldsymbol{y}_{i,n}$ or $\hat{\boldsymbol{y}}_{j,n|n-1}$ can only be used to match once. $\textbf{\textit{A}}_n$ is the assignment matrix, where $A_n(i,j)=1$ if $\boldsymbol{y}_{i,n}$ is assigned to $\hat{\boldsymbol{y}}_{j,n|n-1}$, otherwise $A_n(i,j)=0$. The above optimization problem can be solved using standard solvers \cite{VBO}.

By doing so, based on the matching results of P-ID pairs and the association between GUTI and P-ID after IA, the GUTI are successfully matched with P-IDs during the subsequent beam tracking process.

\section{Performance analysis}

\subsection{Beam accuracy}
For clarification, we compare the frame structure of the conventional beam training/tracking methods, the classic ISAC, and our proposed solution in \textbf{Fig. \ref{Frame_Structure}}. For the beam training/tracking methods, there are no beam association errors since the periodical feedback brings the GUTI information. However, the downlink pilots and uplink feedback are both indispensable, resulting in large overhead and time consumption. In addition, the angles information learned at the latest epoch may be outdated, resulting in that beams may not align owing to the time-varying channel states. The classic ISAC solution removes the pilot overhead and the frequent feedback. Whereas, unless periodically introducing the GUTI feedback, it will still suffer from beam association error since the GUTI information is not contained in the reflected echo. Our proposed solution inherited the advantages of the classic ISAC scheme, including the low overhead and high efficiency. When the network keeps working in CM, the whole downlink block becomes a dual function since it plays both as radar sensing signals and communication data symbols. Furthermore, after obtaining the GUTI information in the initial stage, the BS will subsequently track the UAVs and associate their GUTIs with the P-IDs extracted from the echoes. As a result, the dedicated resources reserved for GUTI feedback can all be saved for transmitting useful data without any beam mismatching.

\subsection{IA delays}	
In addition, we compare the IA delays (namely the times taken by the PSS and RA stages) of the PI-based, exhaustive-search-based, and iterative-search-based procedures. Recall that we have $S_1\ll|\mathcal {S}_{\mathrm{B}}|$ and $S_2\ll|\mathcal {S}_{\mathrm{U}}|$, which makes the IA delay of the proposed solution much shorter than that of the exhaustive search. Even though the PI-based methods remove the ``one-side" delay, they still inevitably suffer from the long search time on the other side. The iterative-search-based procedure causes a disparity between the range of detection and service and faces a long IA delay. In contrast to these procedures, our proposed solution fully utilizes the sensing information of the BS and the UAVs. When the network switches from IM to CM, the beam training is accelerated with the assistance of camera sensing, namely searching in two potential beam sets. The BS's beam is not simply steered toward the UAV but selected from a potential beam set. Once it is properly built, the near-optimal beam could be determined with a short IA delay.

\subsection{Computational Complexity}
In the IA state, since the beam steering is employed with the analog phase shifts and the advanced camera sensing technique could realize a detection speed of more than a hundred frames per second, they are of negligible computational complexity. When considering $T_p$ as a unit time, the IA complexity of our proposed IA procedure can be regarded as $O(|\tilde{\mathcal{S}}_{\mathrm{B}}||\tilde{\mathcal{S}}_{\mathrm{U}}|)$.

In the beam prediction and tracking stage, the EKF requires the execution matrix inversion, having a cubic complexity order of the state vector dimension $V$, i.e., $O(V^3)$. The computational complexity of P-ID generation lies in the dynamic similarity calculation, which is $O(MK^2)$. In the identity matching stage, the KM algorithm has the order of complexity $O(K^3)$ to minimize the globe cost. Please note that $V$, $M$, $|\tilde{\mathcal{S}}_{\mathrm{B}}|$ and $|\tilde{\mathcal{S}}_{\mathrm{U}}|$ are significantly smaller than $K$ in the large-scale UAV network, so the order of computational complexity of the proposed solution can be thus regarded as that of the identity matching stage.

\section{PERFORMANCE EVALUATION AND DISCUSSION}
\subsection{Real-World Experiments} \label{real_world_results}
In this subsection, we conduct some real-world experiments to evaluate the accuracy of camera sensing and the delay of the ISAC-based beam alignment.
\begin{figure}
	\centering
	\subfloat[DJI M300 RTK]{\includegraphics[width=0.4\columnwidth]{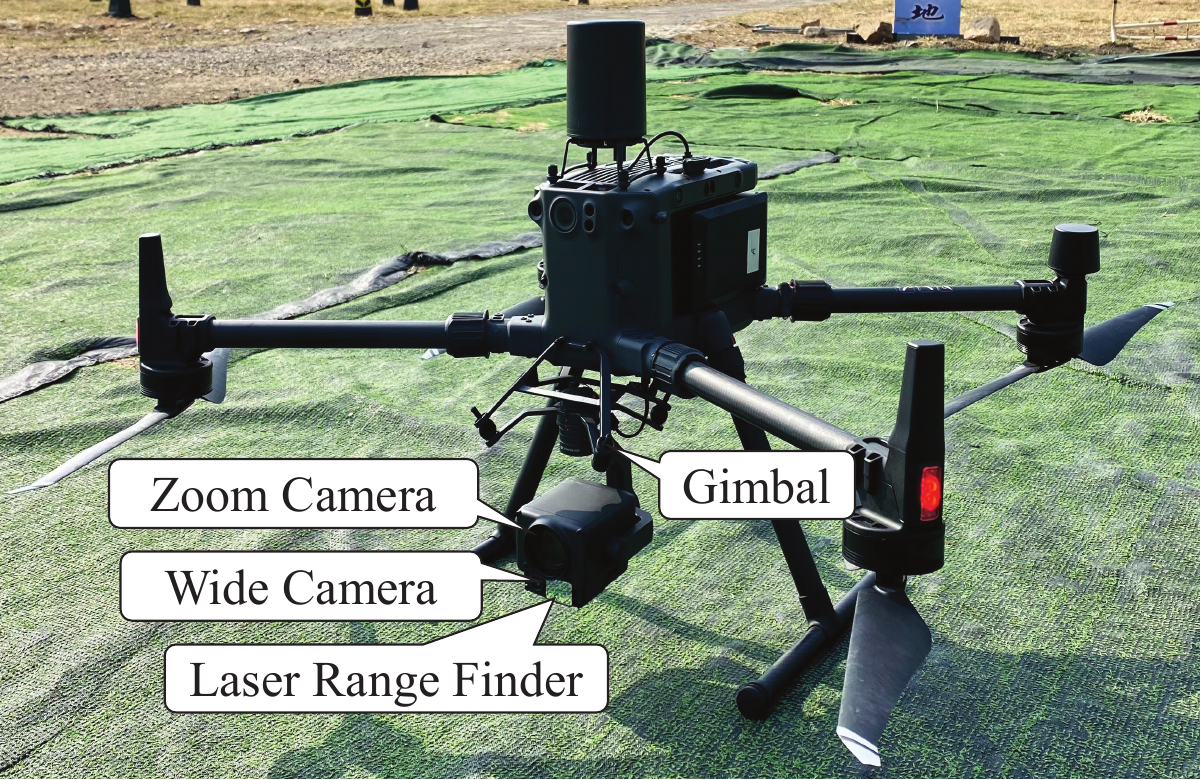}}\quad
	\subfloat[Detection result of ZF-F1200]{\includegraphics[width=0.52\columnwidth]{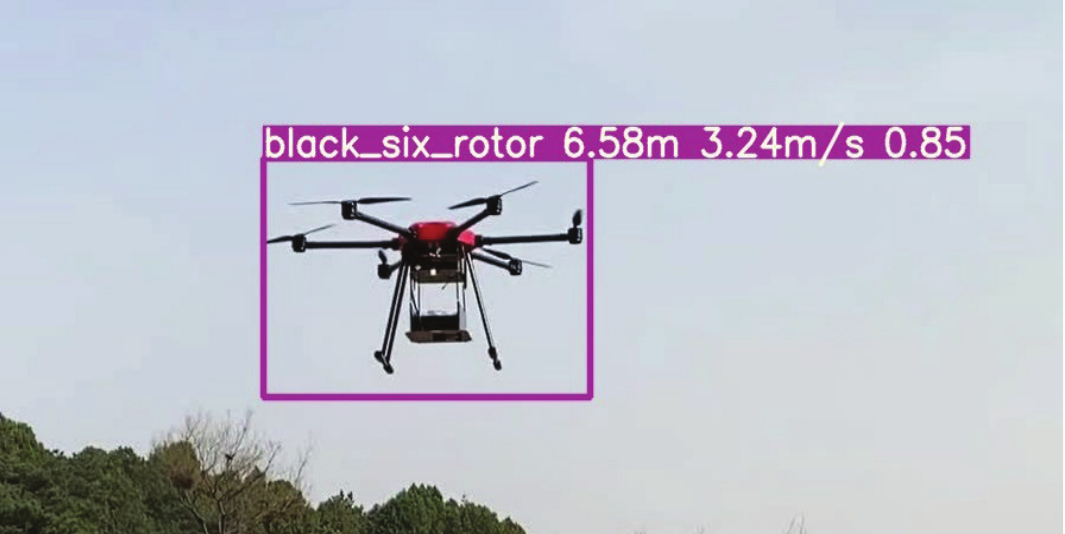}}
	\caption{UAV-based testbed for real-world experiments. } \label{Real_Experiments}
\end{figure}
\begin{table}
	\centering
	\includegraphics[width=1\linewidth]{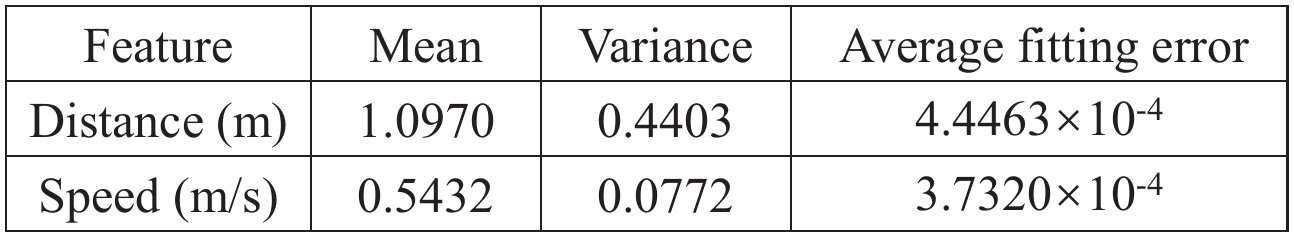}
	\caption{The parameters of Gaussian distribution that fits the results of the real-world experiment.}
	\label{videofittingerror}
\end{table}

As shown in \textbf{Fig. \ref{Real_Experiments}(a)}, we deploy the Zenmuse H20 gimbal \& camera system and LRF system on DJI Matrice 300 (M300) RTK. To build a camera sensing method, we applied YOLO version 5 on M300 to detect the relative distance between it and ZF-F1200, a six-rotor UAV shown in \textbf{Fig. \ref{Real_Experiments}(b)}. M300 and ZF-F1200 initially hover at a height of 15m and 10m respectively, and ZF-F1200 is 8m horizontally behind M300. During the next 10s, they move towards the positive direction of M300 at 2m/s and 2.5m/s and ascend vertically at 2m/s and 2.5m/s, respectively. The above flight is repeated 10 times with different initial locations. We calculate the relative velocity and distance based on the exported data of positioning and speed sensors and regard them as the baseline. \textbf{Table \ref{videofittingerror}} shows that the sensing errors follow Gaussian distributions, and the experimental results can be fitted with an average probability error below $4.5\times10^{-4}$.
\begin{figure}
	\centering
	\subfloat[Transmitter]{\includegraphics[width=0.46\columnwidth]{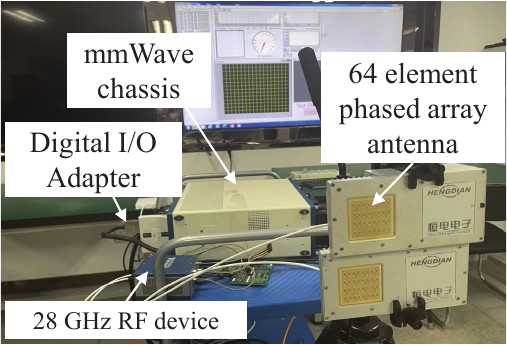}}\quad
	\subfloat[Receiver (moving target)]{\includegraphics[width=0.46\columnwidth]{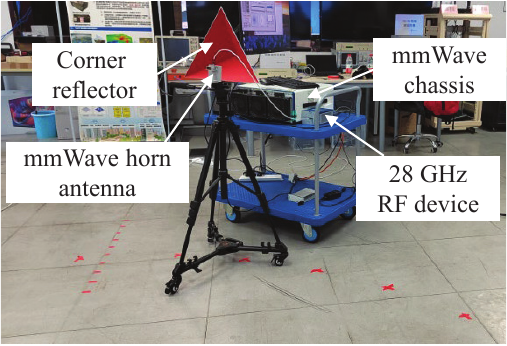}}
	\caption{Real-world experiment environment: the hardware testbed for time-division ISAC system.} \label{Testbed}
\end{figure}

\begin{figure}
	\centering
	\includegraphics[width=1\linewidth]{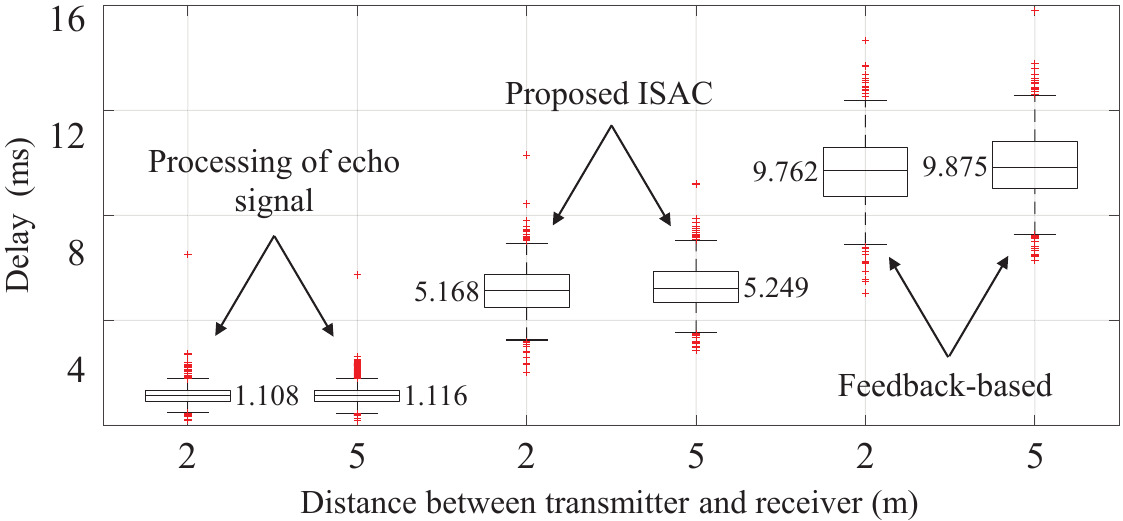}
	\caption{Performance comparison of time consumption of echo signal processing, the proposed ISAC solution, and the feedback-based solution.}
	\label{Timeconsumption}
\end{figure}

We also evaluate the time consumption with the hardware testbed shown in \textbf{Fig. \ref{Testbed}}, which has been established in our recent work \cite{testbed}. The specific hardware design will not be introduced here due to page limitations, and readers can refer to \cite{testbed} for more details. As shown in \textbf{Fig. \ref{Timeconsumption}}, compared with the feedback-based one, the proposed ISAC solution reduces the total delay by 4.594 ms and 4.626 ms in 2 m and 5 m, respectively. The unique echo processing delay only takes about 1.1 ms. In addition, the matching result of 10 nodes with two physical characteristics can be obtained within 0.5 ms on average, and this is the only additional time consumption when compared with the classic ISAC method. Although the experiment is performed indoors rather than in an aerial communication scene, it is sufficient to make conclusions that our proposed ISAC approach outperforms the feedback-based solution in terms of time consumption and is not much inferior to the classic ISAC.

\subsection{numerical results}
In this subsection, we present the numerical results to validate the effectiveness of the proposed DIA solution. Let's consider a network with 10 UAVs moving freely in 3-D spaces. Their initial positions are randomly generated on a hemispherical surface with a radius of 100 m, and the BS is located at the center of the sphere. All UAVs fly towards the BS in the horizontal direction and randomly deviate within 20 degrees and a vertical direction that is randomly distributed within $\pm$10 degrees. The speed is limited to 18$\sim$20 m/s. The BS is operating in $f_c=28$ GHz \cite{28GHz}. The duration for each time slot is $\Delta T=0.02s$. Other parameters are set as $G=10$, $\sigma=\sigma_r=1$, $\tilde{\alpha}=1$, $\sigma_{p(i)}=0.02$ m, $\sigma_{v(i)}=0.2$ m/s, $a_1=6.7\times10^{-7}$, $a_2=2\times10^4$, $a_3 = 1$, $A_1 = -0.4568$, $A_2 = 0.0470$, $A_3 = -0.63$, $A_4 = 1.63$, ${\rm K_R}_{\rm min} = 0$ dB, ${\rm K_R}_{\rm max} = 30$ dB. We run 2000 Monte Carlo trials to evaluate the performance, and the average results are discussed below.

\begin{remark}
	Unless otherwise specified, the EKF-ISAC refers to the proposed predictive ISAC solution for a single UAV, while the DIA-based ISAC refers to the proposed predictive ISAC solution for multi-UAV scenarios.
\end{remark}

\subsubsection{Matching accuracy}

In \textbf{Table \ref{Table_Matching}}, we compare the matching accuracy (the proportion of beams that correctly match the intended UAVs) for different solutions.

The position and velocity-based solution only utilize the position and speed to calculate the matching cost $D_n$. The static weight-based solution gives equal weight, namely 0.5, to both the position and velocity, and the proposed dynamic weight-based solution assigns weights according to the feature's time-varying prevalence. We also compare different similarity metrics, including the Euclidean distance (ED), Mahalanobis distance (MD), and the Wasserstein distance, which is also known as the Earth Mover’s distance (EMD). In addition, the performance from various matching algorithms, e.g., the Kuhn-Munkres (KM) algorithm, the function ``matchpairs'' that MATLAB provided, the Greedy algorithm, the Auction algorithm, the LAPJV, and the LAPJV-Fast algorithm.
\begin{table}
	\centering
	\includegraphics[width=0.9\linewidth]{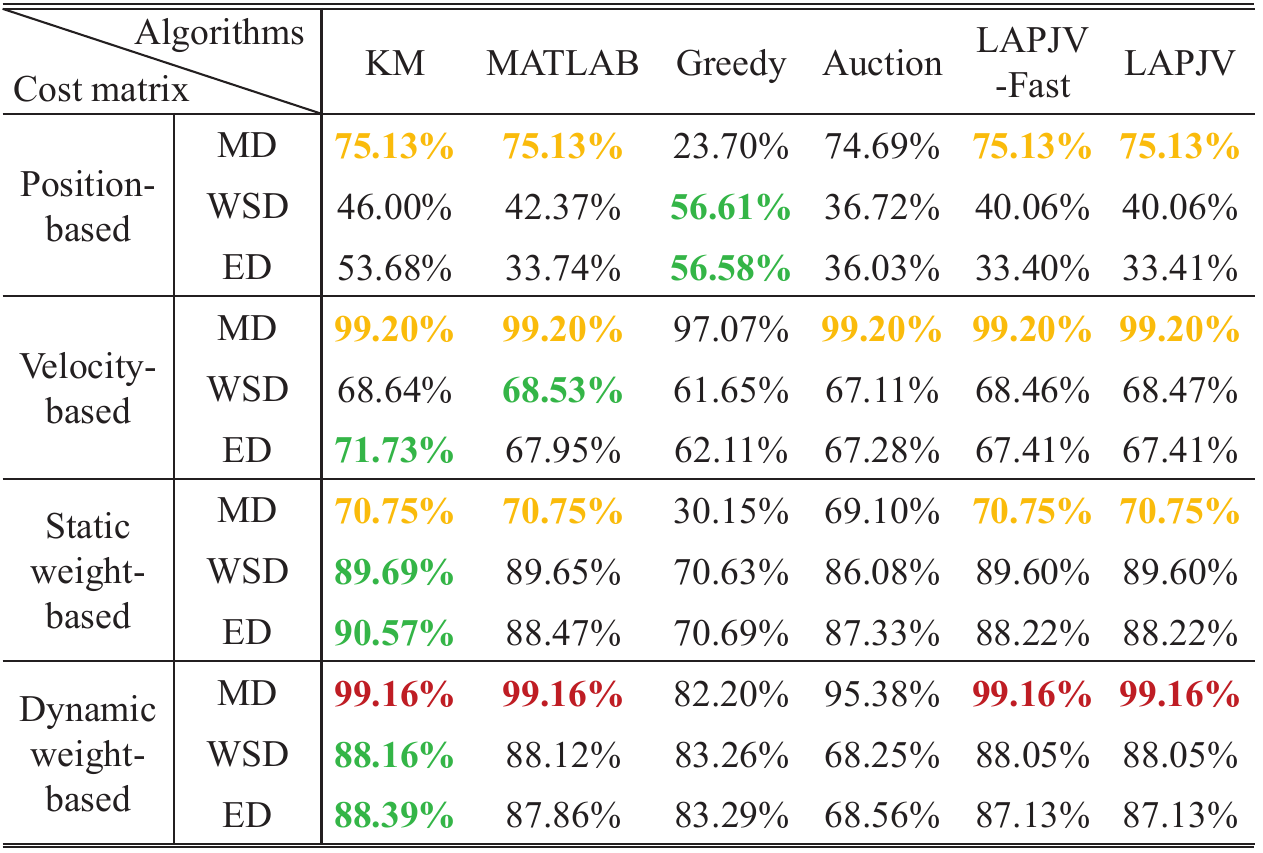}
	\caption{Matching accuracy of different distance metrics, matching algorithms, and calculation methods for cost matrix. The top accuracy results are marked as red, the best results for each distance metric are marked as blue, and the tied best results are marked as yellow.}
	\label{Table_Matching}
\end{table}

\begin{figure}
	\centering
	\includegraphics[width=0.9\linewidth]{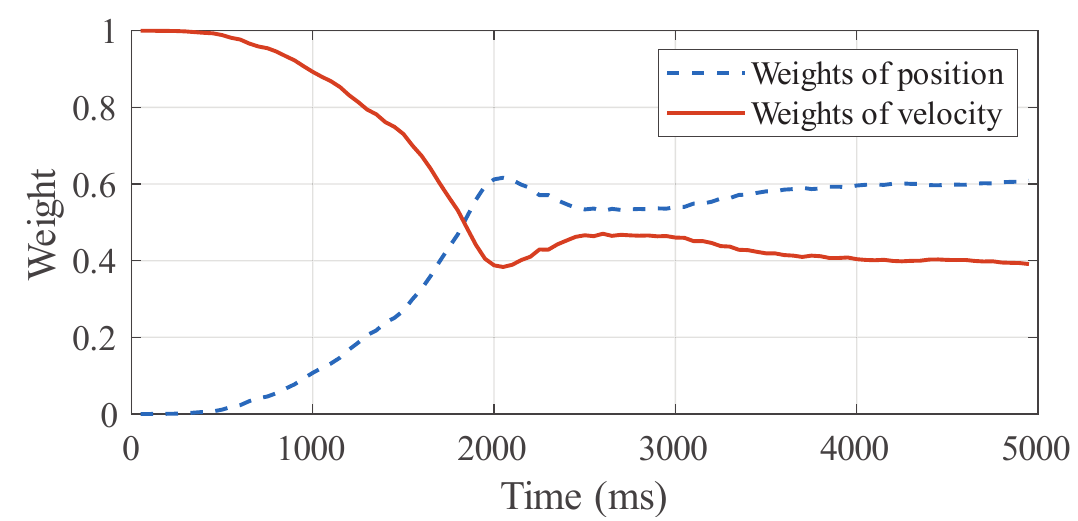}
	\caption{The weight for location and velocity of multiple UAVs during the whole flight.}
	\label{weight}
\end{figure}

The position-based solution has the worst performance in most cases since UAVs tend to occasionally approach each other when moving over the BS. The velocity-based one performs better than the position-based one since UAV's velocities are distinguishable in most cases. Although the static weight-based solution utilizes both position and velocity to distinguish UAVs, it still suffers from some errors since similar positions still take half of the weight, which induces a relatively reduced effect of velocity in the cost matrix. Our proposed dynamic weight-based solution attains a higher matching accuracy since the UAVs' P-IDs are generated according to their physical characteristic's prevalence. That is, the location information plays a more significant role in the matching matrix once the velocities are similar, and vice versa. It can be seen from \textbf{Fig. \ref{weight}} that the velocity has a large weight during 0ms$\sim$1800 ms since the positions of all UAVs are more similar than the velocities. However, after 1800ms, UAVs have already flown past the BS and there tend to be different positions. Therefore, the prevalence of position continues to decline, so it plays a more significant role in distinguishing UAVs. There will be a large weight for positions, which will dominate the matching matrix. Nevertheless, this performance advantage does not appear in all distance indicators owing to the various performance of distance metrics, which are discussed as follows.

\begin{table}
	\centering
	\includegraphics[width=0.9\linewidth]{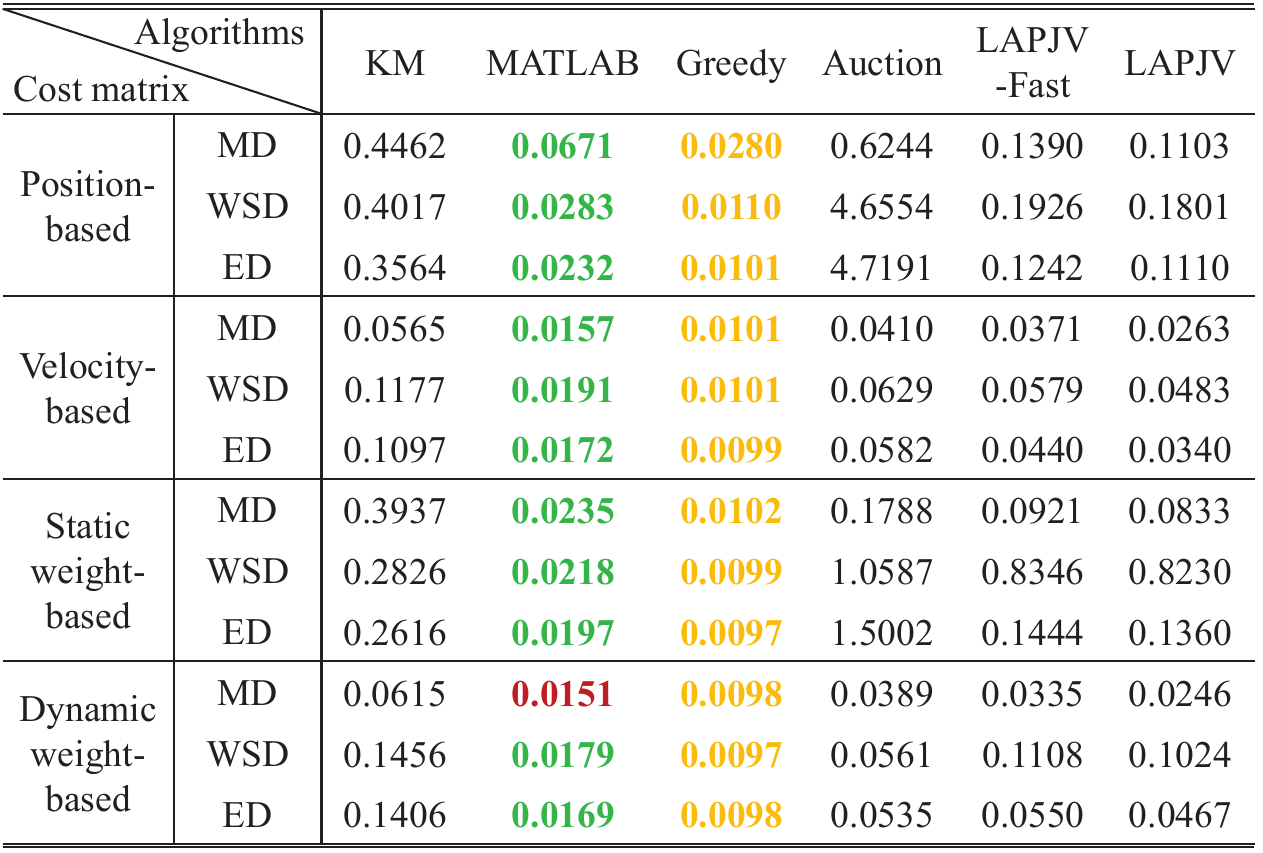}
	\caption{Time consumption (ms) of different distance metrics, matching algorithms, and calculation methods for cost matrix. The minimum time consumption results for each distance metric are marked as yellow. The second-best results are marked as green, among which the best one is marked as red.}
	\label{Table_Time}
\end{table}

For the ED, if there are significant differences in the scales of values, the components with high variability will play a decisive role in the sum of squares of Euclidean distances, while the components with low variability will have little effect. For example, when calculating similarity using the UAVs' position, the z-coordinate (height) of drones is limited to 20 to 30 meters, but the x-coordinate and y-coordinate are changed within -100 to 100 meters, the impact of height on similarity will be seriously ignored. The same reason also leads to poor performance in the velocity-based solution. Although the WSD-based solution also has an excellent performance when utilizing the static-weight-based solution, it can not accurately distinguish UAVs with one physical feature, whether using positions and velocities. The above defects can be corrected by introducing covariance, namely utilizing the MD, which characterizes the covariance distance of the UAV's state vector. It can be seen from Table \ref{Table_Matching} that the MD has the best performance in both the single feature-based solutions. The reason is that the relationship between various features can be considered and it is independent of the scale of measurement since it is not affected by dimensions. In addition, it outperforms the ED- and WSD-based solutions in when using the dynamic weights, indicating a better performance when paired with the method proposed in Sec. \ref{Prevalence}.

For the given cost matrix, the performance difference between algorithms also differs much. The Greedy algorithm follows the principle of ``first come, first served'', which prioritizes the allocation of low-cost matching to the UAVs that were earlier assigned, resulting in the wrong assignment to the UAVs that were later assigned. Even if it has the minimum time consumption, as shown in Table \ref{Table_Time}, and performs well in some position-based solutions, it is futile because this feature cannot be used to accurately distinguish UAVs. For the auction algorithm, the unreasonable value of Epsilon makes it easy to miss the optimal matching sometime and suffer the maximum time consumption. For the KM, MATLAB, LAPJV, and LAPJV-Fast algorithms, the accuracy difference under dynamic features is very small, and all of them almost reach 100\% matching accuracy when utilizing the proposed dynamic weight-based MD solution. Since the function ``matchpairs'' provided by MATLAB has the minimum time cost, it will be used to solve the assignment problem in the following experiments.

Note that there are still cases where UAVs are indistinguishable by any solution since the optimal accuracy in Table \ref{Table_Matching} only achieves 99.16\%. However, an average gap of about 0.84\% from perfect matching is still acceptable owing to the fact that we only utilize two physical characteristics, and the experimental results are sufficient to prove the effectiveness of the dynamic weight and MD. In the future, the full utilization of all available physical characteristics (e.g., the micro-Doppler characteristics) is expected to approach the perfect matching performance since it is extremely rare that all physical characteristics are simultaneously indistinguishable.
\subsubsection{Tracking errors}
\begin{figure}
	\centering
	\includegraphics[width=0.9\linewidth]{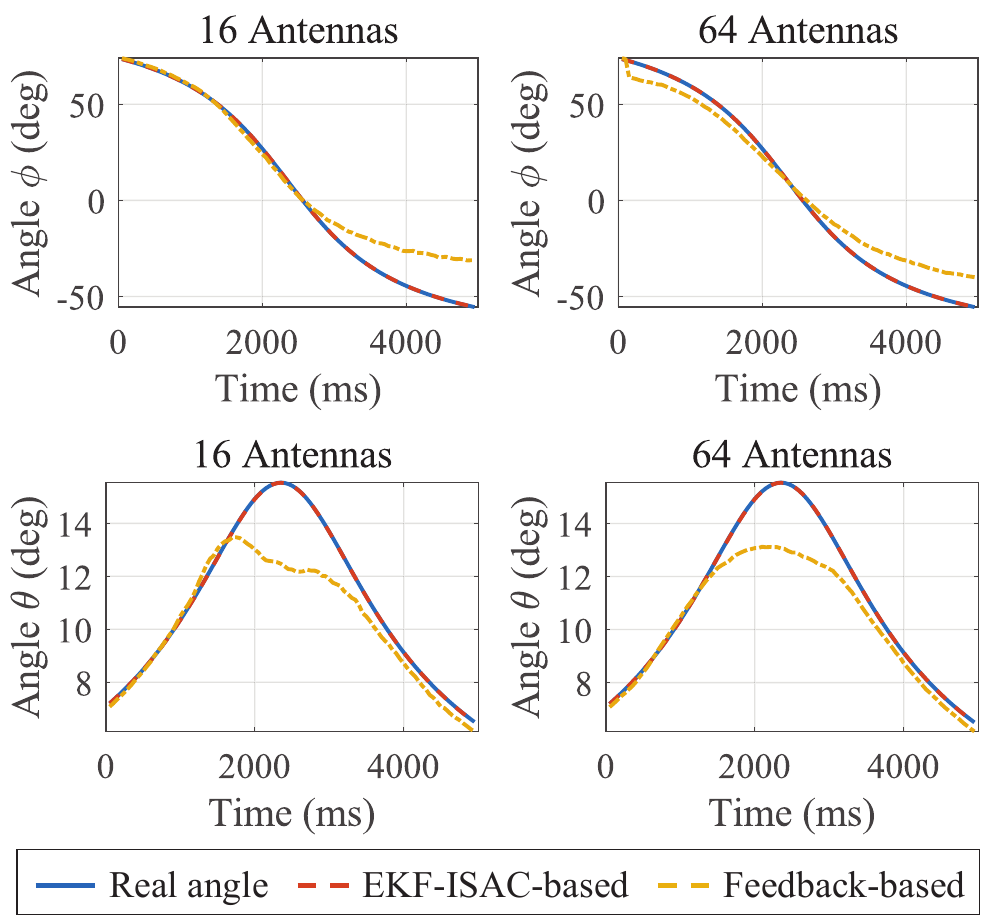}
	\caption{Angle tracking performances: EKF-ISAC and feedback schemes for single UAV.}
	\label{Angle_all}
\end{figure}
In \textbf{Fig. \ref{Angle_all}}, we show the prediction performances for the feedback-based solution and the proposed EKF-based ISAC solution.

It should be noted that the beam association errors are not considered in \textbf{Fig. \ref{Angle_all}}. Recall that UAVs will fly over the top of the BS at about 1800 ms, so the angles $\phi$ and $\theta$ change faster during this period. The prediction accuracy of the feedback-based solution drops owing to the following reasons: i) there is only one single pilot being exploited for tracking and ii) it requires a receive beamformer to combine the pilot signal and inevitably leads to the loss of the detailed angle information. Compared with the EKF-ISAC solution, the angles estimated by the feedback-based one have large errors. The EKF-ISAC solution utilizes the whole echo signal block for sensing, and the matched filter gain is 10 times of that in the feedback-based one. As a result, it has a better tracking performance than the feedback-based one.

\subsubsection{Achievable rates}
\begin{figure}
	\centering
	\includegraphics[width=0.9\linewidth]{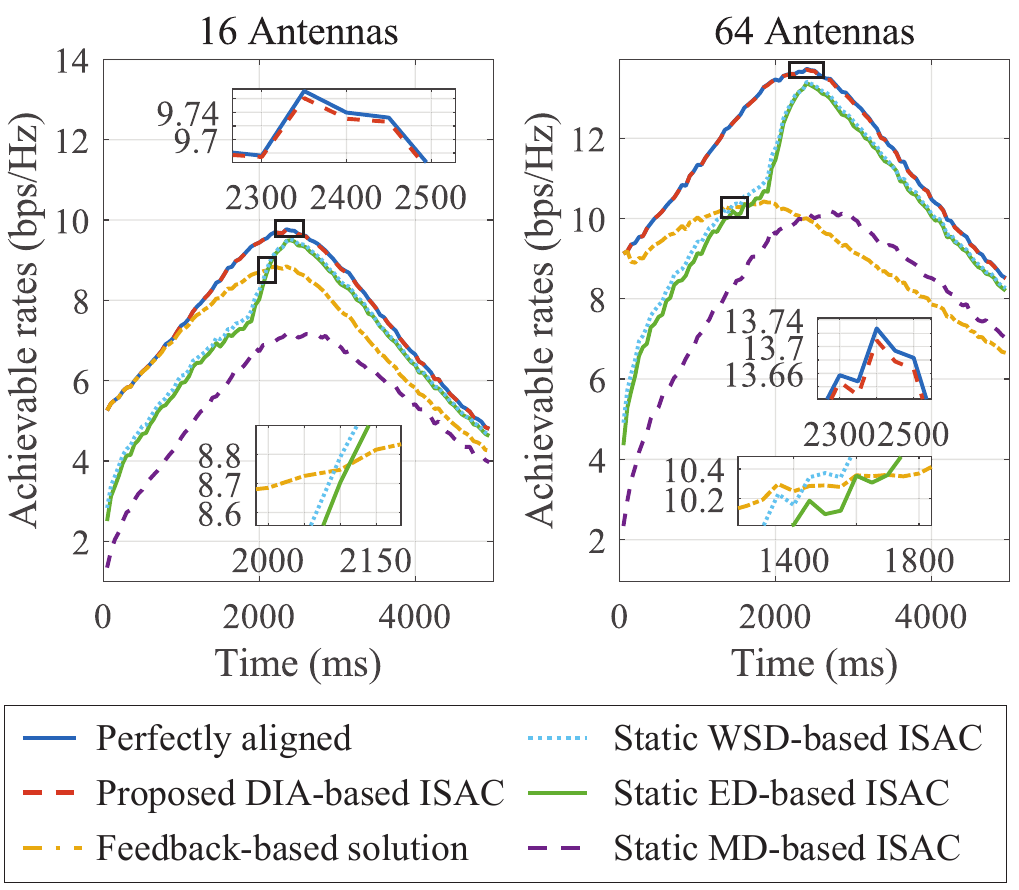}
	\caption{Average achievable rates of different schemes.}
	\label{Achievable_Rate}
\end{figure}
In \textbf{Fig. \ref{Achievable_Rate}}, we compare the average achievable rates of six solutions, where the results that the predicted angles perfectly match the real angle are denoted as the rates that ``perfectly aligned''. Please note that the achievable rate of a link will be zero if the P-ID of a UAV is not correctly associated with its GUTI, i.e., a beam is wrongly associated with an unintended UAV.

The proposed DIA-based ISAC solution maintains the highest rates since it has the smallest tracking error and almost no matching errors. The feedback-based scheme notifies the BS of the UAV's GUTI in real-time, so there is no beam failure problem caused by mismatch, and the rates closely follow our solution at the beginning when the angle variation is relatively slow. Subsequently, the rate of the feedback-based scheme drops when the UAVs are approaching the BS, which is consistent with the associated angle tracking performance in \textbf{Fig. \ref{Angle_all}}. In addition, it decreases drastically in the 64-antenna scenario given the narrower beam and higher misalignment probability. When compared with our proposed solution, it averagely suffers 7.95\% and 20.34\% of the rates loss in the 16- and 64-antenna cases, respectively.

\begin{figure}
	\centering
	\subfloat[IA delay versus the number of bits that control the BS's phase. ]{\includegraphics[width=0.9\columnwidth]{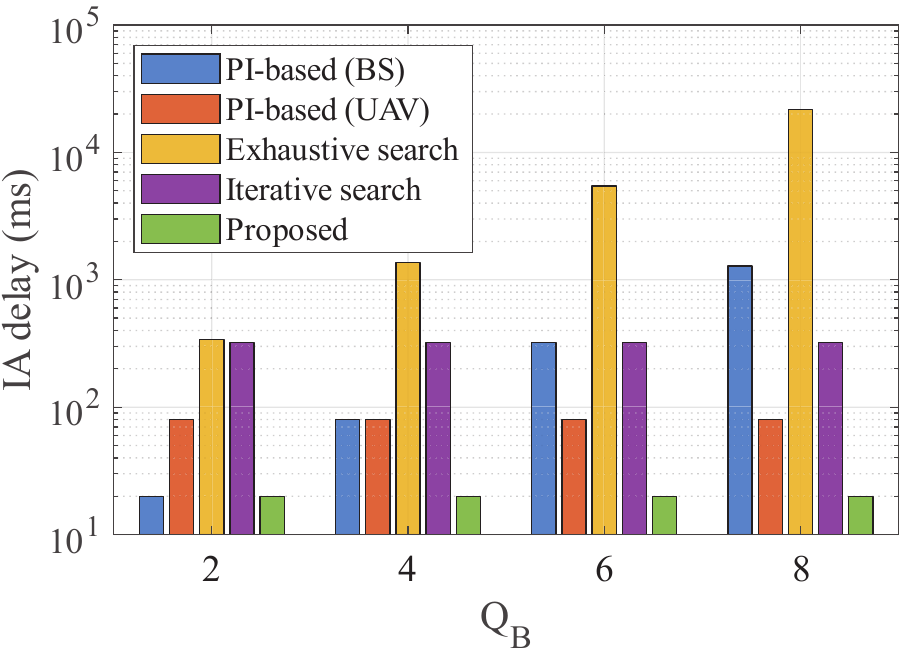}}\\
	\subfloat[IA delay versus the number of bits that control the UAV's phase. ]{\includegraphics[width=0.9\columnwidth]{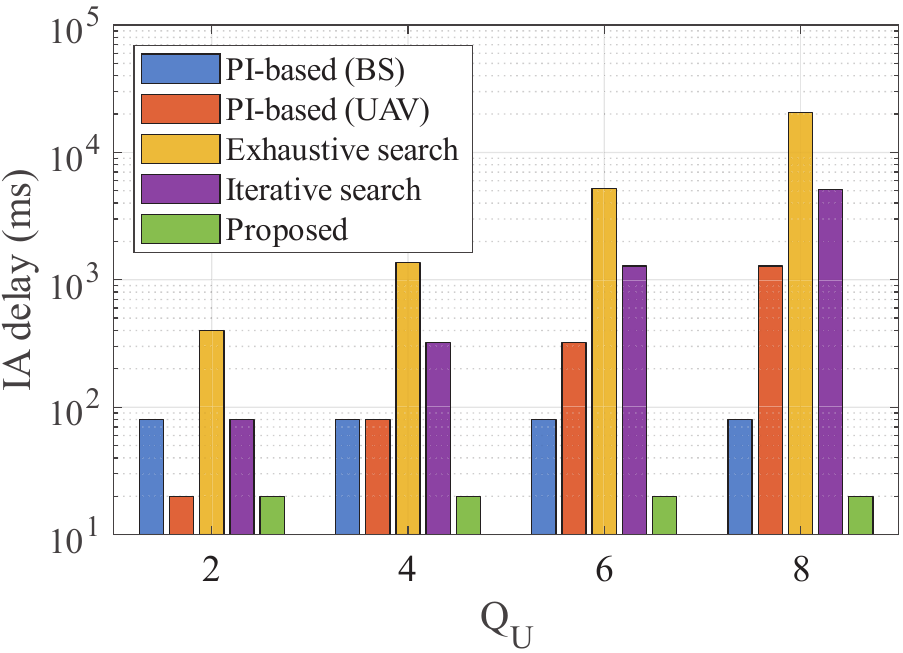}}
	\caption{Comparison of IA delay for different schemes.} \label{IA-delay-experiment}
\end{figure}
The static WSD-, ED-, and MD-based ISAC calculate the similarity based on three distance metrics and assign equal weight to position and velocity. It should be noted that both of the first two solutions have similar rates during the whole flight owing to the similar accuracy shown in rows 8 and 9 of the Table \ref{Table_Matching}. Besides, in the beginning, the rates of them drop to about half of the optimal rates; and after 1800 ms, both of them continue to improve until approach our solution. This is because the positions of UAVs are too similar in the beginning and gradually become distinguishable after 1800 ms. In other words, the performance of both the MD and WSD are sensitive to similar positions since the positions with large scale and high variability play a more decisive role than the velocity in calculating similarity. For the ED-based one, it results in 11.73\% and 11.63\% degradation of average rates in the 16-antenna and 64-antenna cases, respectively. Even so, they are still better than the static MD-based ISAC, which suffers from larger matching errors even when the position becomes distinguishable after 1800 ms, because it transforms the night elements of the UAV's state to the same scale according to their covariance, making it more difficult to distinguish similar features with static weights. According to statistics, the rates of the static MD-based ISAC averagely decrease to about 70.67\% and 70.69\% of our solution in the 16- and 64-antenna cases, respectively.

\subsubsection{IA delay}
In Fig. \ref{IA-delay-experiment}, the IA delays of five schemes are compared. The PI-based schemes utilize the prior information of UAV and BS, respectively. The exhaustive search traverses all possible beam directions to find the one with the highest SNR. The iterative search employs hierarchical beam codebooks. Our proposed sensing-assisted IA scheme utilizes the camera sensing ability at both the BS side and the UAV side. The cardinality $S_1$ and $S_2$ are both set to 2, which means the sensing information helps to determine two potential beam directions. The $Q_U$ in Fig. \ref{IA-delay-experiment}(a) and the $Q_B$ in Fig. \ref{IA-delay-experiment}(b) are all set to 4. According to the delay parameters in \cite{Position-aided-beam-learning}, $T_p$ is set as 5ms. That is, the BS sends a PSS once every 5ms during the PSS detection process, and after that, the UAV transmits an RA preamble once every 5ms.

The delay of the exhaustive search increases with the increase of $Q_U$ and $Q_B$ since it requires $(|S_B|+1)\times|S_U|\times T_p$ scan slots at the worst to complete the IA process. The delay of the iterative search will not be influenced by $Q_B$ since here we implement the scheme in \cite{interactive-search}. That is, the BS obtains the prior information of the UAV, and the iterative search is only performed on the UAV side. However, it still suffers from a large delay when $Q_U$ becomes larger. Interestingly, its delay under different $Q_U$ is close to one-quarter of the exhaustive method, which is owing to the fact that the 16 exhaustive searches are replaced by 4 hierarchical searches. For the two PI-based methods, the beam scan at the UAV (BS) side is no longer needed due to the introduction of BS (UAV) location information. The PI-based (BS) and PI-based (UAV) methods need $|S_B|\times T_p$ and $|S_U|\times T_p$ scan slots to complete the IA process. Our proposed scheme reduces the scanning space of the UAV and the BS to $S_1$ and $S_2$, respectively. That is, only $S_1\times S_2\times T_p$ scan slots are needed in the worst case, so it has the lowest delay among all schemes and is obviously superior to them when $Q_B$ and $Q_U$ become larger.

\section{conclusion}
In this paper, the proposed sensing-assisted beam management solution utilizes the VD information to improve the performance of delay and accuracy of beam alignment. The sensing under occlusion and systematical evaluation of beam alignment in actual scenarios will be designated as our future work.

\section*{ACKNOWLEDGEMENT}
\label{ACKNOWLEDGEMENT}
This work was partly supported by the Major Research Projects of the National Natural Science Foundation of China (92267202), the National Key Research and Development Project (2020YFA0711303), and the BUPT Excellent Ph.D. Students Foundation (CX2022208).

\bibliographystyle{gbt7714-numerical}
\bibliography{myref_new}

\begin{thebibliography}{25}
\providecommand{\natexlab}[1]{#1}
\providecommand{\url}[1]{#1}
\expandafter\ifx\csname urlstyle\endcsname\relax\else
  \urlstyle{same}\fi
\expandafter\ifx\csname href\endcsname\relax
  \DeclareUrlCommand\doi{\urlstyle{rm}}
  \def\eprint#1#2{#2}
\else
  \def\doi#1{\href{https://doi.org/#1}{\nolinkurl{#1}}}
  \let\eprint\href
\fi

\bibitem[Cui et~al.(2022)Cui, Zhang, Feng, Wei, Shi, and Yang]{cui2022topology}
CUI Y, ZHANG Q, FENG Z, et~al.
\newblock Topology-aware resilient routing protocol for {FANET}s: An adaptive
  {Q}-learning approach\allowbreak[J].
\newblock IEEE Internet of Things Journal, 2022, 9\allowbreak (19):
  18632-18649.

\bibitem[Zhang et~al.(2022)Zhang, Yang, Feng, Cui, Dai, Qin, Li, and
  Zhang]{on-purpose}
ZHANG P, YANG H, FENG Z, et~al.
\newblock Towards intelligent and efficient 6{G} networks: {JCSC} enabled
  on-purpose machine communications\allowbreak[J].
\newblock IEEE Wireless Communications, 2022.

\bibitem[Xiao et~al.(2021)Xiao, Zhu, Liu, Yi, Zhang, Xia, and
  Schober]{millimeter-wave-UAV-survey}
XIAO Z, ZHU L, LIU Y, et~al.
\newblock A survey on millimeter-wave beamforming enabled uav communications
  and networking\allowbreak[J].
\newblock IEEE Communications Surveys \& Tutorials, 2021, 24\allowbreak (1):
  557-610.

\bibitem[Shah et~al.(2021)Shah, Aditya, and Rangan]{DRX}
SHAH S~H~A, ADITYA S, RANGAN S.
\newblock Power-efficient beam tracking during connected mode drx in mmwave and
  sub-thz systems\allowbreak[J].
\newblock IEEE Journal on Selected Areas in Communications, 2021, 39\allowbreak
  (6): 1711-1724.

\bibitem[Giordani et~al.(2019)Giordani, Polese, Roy, Castor, and
  Zorzi]{standalone-magazine}
GIORDANI M, POLESE M, ROY A, et~al.
\newblock Standalone and non-standalone beam management for 3gpp nr at
  mmwaves\allowbreak[J].
\newblock IEEE Communications Magazine, 2019, 57\allowbreak (4): 123-129.

\bibitem[Chiu et~al.(2019)Chiu, Ronquillo, and Javidi]{interactive-search}
CHIU S~E, RONQUILLO N, JAVIDI T.
\newblock Active learning and {CSI} acquisition for mmwave initial
  alignment\allowbreak[J].
\newblock IEEE Journal on Selected Areas in Communications, 2019, 37\allowbreak
  (11): 2474-2489.

\bibitem[Hu et~al.(2020)Hu and He]{Position-aided-beam-learning}
HU A, HE J.
\newblock Position-aided beam learning for initial access in mmwave mimo
  cellular networks\allowbreak[J].
\newblock IEEE Systems Journal, 2020, 16\allowbreak (1): 1103-1113.

\bibitem[Yan et~al.(2020)Yan, Fang, Hao, and Fang]{dual-band-search}
YAN L, FANG X, HAO L, et~al.
\newblock A fast beam alignment scheme for dual-band hsr wireless
  networks\allowbreak[J].
\newblock IEEE Transactions on Vehicular Technology, 2020, 69\allowbreak (4):
  3968-3979.

\bibitem[Cui et~al.(2022)Cui, Zhang, Feng, Wei, Shi, Fan, and
  Zhang]{Dual_Identity}
CUI Y, ZHANG Q, FENG Z, et~al.
\newblock Dual identities enabled low-latency visual networking for {UAV}
  emergency communication\allowbreak[C]//\allowbreak
GLOBECOM 2022-2022 IEEE Global Communications Conference.
\newblock IEEE, 2022: 474-479.

\bibitem[Cui et~al.(2023)Cui, Feng, Zhang, Wei, Xu, and Zhang]{Dual_ID}
CUI Y, FENG Z, ZHANG Q, et~al.
\newblock Toward trusted and swift uav communication: Isac-enabled dual
  identity mapping\allowbreak[J/OL].
\newblock IEEE Wireless Communications, 2023, 30\allowbreak (1): 58-66.
\newblock DOI: \doi{10.1109/MWC.003.2200207}.

\bibitem[Liu et~al.(2020)Liu, Yuan, Wei, Liu, and
  Ng]{UAV_Learning_Beamforming1}
LIU C, YUAN W, WEI Z, et~al.
\newblock Location-aware predictive beamforming for {UAV} communications: A
  deep learning approach\allowbreak[J].
\newblock IEEE Wireless Communications Letters, 2020, 10\allowbreak (3):
  668-672.

\bibitem[Yuan et~al.(2020)Yuan, Liu, Liu, Li, and
  Ng]{UAV_Learning_Beamforming2}
YUAN W, LIU C, LIU F, et~al.
\newblock Learning-based predictive beamforming for {UAV} communications with
  jittering\allowbreak[J].
\newblock IEEE Wireless Communications Letters, 2020, 9\allowbreak (11):
  1970-1974.

\bibitem[Lyu et~al.(2022)Lyu, Zhu, and Xu]{UAV_ISAC1}
LYU Z, ZHU G, XU J.
\newblock Joint trajectory and beamforming design for {UAV}-enabled integrated
  sensing and communication\allowbreak[C]//\allowbreak
ICC 2022-IEEE International Conference on Communications.
\newblock IEEE, 2022: 1593-1598.

\bibitem[Wei et~al.(2022)Wei, Liu, Ng, and Schober]{UAV_ISAC3}
WEI Z, LIU F, NG D~W~K, et~al.
\newblock Safeguarding {UAV} networks through integrated sensing, jamming, and
  communications\allowbreak[C]//\allowbreak
ICASSP 2022-2022 IEEE International Conference on Acoustics, Speech and Signal
  Processing (ICASSP).
\newblock IEEE, 2022: 8737-8741.

\bibitem[Cui et~al.(2023)Cui, Zhang, Feng, Liu, Shi, Fan, and
  Zhang]{cui2023specific}
CUI Y, ZHANG Q, FENG Z, et~al.
\newblock Specific beamforming for multi-uav networks: A dual identity-based
  isac approach\allowbreak[A].
\newblock 2023.

\bibitem[Liu et~al.(2020)Liu, Yuan, Masouros, and Yuan]{Com_Served_By_Sen}
LIU F, YUAN W, MASOUROS C, et~al.
\newblock Radar-assisted predictive beamforming for vehicular links:
  Communication served by sensing\allowbreak[J].
\newblock IEEE Transactions on Wireless Communications, 2020, 19\allowbreak
  (11): 7704-7719.

\bibitem[Hilario~Re et~al.(2020)Hilario~Re, Podilchak, Rotenberg, Goussetis,
  and Lee]{independent_transmit_and_receive}
HILARIO~RE P~D, PODILCHAK S~K, ROTENBERG S~A, et~al.
\newblock Circularly polarized retrodirective antenna array for wireless power
  transmission\allowbreak[J].
\newblock IEEE Transactions on Antennas and Propagation, 2020, 68\allowbreak
  (4): 2743-2752.

\bibitem[Yuan et~al.(2020)Yuan, Liu, Masouros, Yuan, Ng, and
  Gonz{\'a}lez-Prelcic]{Self}
YUAN W, LIU F, MASOUROS C, et~al.
\newblock Bayesian predictive beamforming for vehicular networks: A
  low-overhead joint radar-communication approach\allowbreak[J].
\newblock IEEE Transactions on Wireless Communications, 2020, 20\allowbreak
  (3): 1442-1456.

\bibitem[Liu et~al.(2021)Liu, Xiong, Lu, Ni, Fan, and
  Letaief]{Channel_Los_Rician}
LIU Y, XIONG K, LU Y, et~al.
\newblock Uav-aided wireless power transfer and data collection in rician
  fading\allowbreak[J].
\newblock IEEE Journal on Selected Areas in Communications, 2021, 39\allowbreak
  (10): 3097-3113.

\bibitem[You et~al.(2020)You and Zhang]{Los_P}
YOU C, ZHANG R.
\newblock Hybrid offline-online design for uav-enabled data harvesting in
  probabilistic los channels\allowbreak[J].
\newblock IEEE Transactions on Wireless Communications, 2020, 19\allowbreak
  (6): 3753-3768.

\bibitem[Ngo(2015)]{mMIMO_Theory}
NGO H~Q.
\newblock Massive {MIMO}: Fundamentals and system designs\allowbreak[M].
\newblock Link{\"o}ping University Electronic Press, 2015.

\bibitem[Kay(1993)]{Variance}
KAY S~M.
\newblock Fundamentals of statistical signal processing: estimation
  theory\allowbreak[M].
\newblock Prentice-Hall, Inc., 1993.

\bibitem[Zhao et~al.(2019)Zhao, Cui, Gao, Guo, and Gao]{VBO}
ZHAO X~Q, CUI Y~P, GAO C~Y, et~al.
\newblock Energy-efficient coverage enhancement strategy for 3-{D} wireless
  sensor networks based on a vampire bat optimizer\allowbreak[J].
\newblock IEEE Internet of Things Journal, 2019, 7\allowbreak (1): 325-338.

\bibitem[Zhang et~al.(2022)Zhang, Sun, Gao, Wang, and Feng]{testbed}
ZHANG Q, SUN H, GAO X, et~al.
\newblock Time-division {ISAC} enabled connected automated vehicles cooperation
  algorithm design and performance evaluation\allowbreak[J].
\newblock IEEE Journal on Selected Areas in Communications, 2022, 40\allowbreak
  (7): 2206-2218.

\bibitem[Wang et~al.(2019)Wang, Phelps, Rupakula, Zihir, and Rebeiz]{28GHz}
WANG Y, PHELPS T, RUPAKULA B, et~al.
\newblock 28 {GH}z 5g-based phased-arrays for {UAV} detection and automotive
  traffic-monitoring radars\allowbreak[C]//\allowbreak
2019 IEEE International Symposium on Phased Array System \& Technology (PAST).
\newblock IEEE, 2019: 1-4.

\end{thebibliography}

\biographies

\begin{CCJNLbiography}{figures/Cui}{Yanpeng Cui}
	received the B.S. degree from the Henan University of Technology, Zhengzhou, China, in 2016, and the M.S. degree from the Xi’an University of Posts and Telecommunications, Xi’an, China, in 2020. He is currently pursuing the Ph.D. degree with the School of Information and Communication Engineering, Beijing University of Posts and Telecommunications (BUPT), Beijing, China. His current research interests include the Flying ad hoc networks, and integrated sensing and communication for UAV networks.
\end{CCJNLbiography}

\begin{CCJNLbiography}{figures/Zhang}{Qixun Zhang}
	(M’12) received the B.E. and the Ph.D. degree from BUPT, Beijing, China, in 2006 and 2011, respectively. From Mar. to Jun. 2006, he was a Visiting Scholar at the University of Maryland, College Park, Maryland. From Nov. 2018 to Nov. 2019, he was a Visiting Scholar in the Electrical and Computer Engineering Department at the University of Houston, Texas. He is a Professor with the Key Laboratory of Universal Wireless Communications, Ministry of Education, and the School of Information and Communication Engineering, BUPT. His research interests include 5G mobile communication system, integrated sensing and communication for autonomous driving vehicle, mmWave communication system, and unmanned aerial vehicles (UAVs) communication. He is active in ITU-R WP5A/5C/5D standards.
\end{CCJNLbiography}

\begin{CCJNLbiography}{figures/Feng}{Zhiyong Feng}
	received her B.S., M.S., and Ph.D. degrees from BUPT, Beijing, China. She is a Professor with the School of Information and Communication Engineering, BUPT, and the director of the Key Laboratory of Universal Wireless Communications, Ministry of Education, China. Her research interests include wireless network architecture design and radio resource management in 5th generation mobile networks (5G), spectrum sensing and dynamic spectrum management in cognitive wireless networks, universal signal detection and identification, and network information theory. She is active in standards development, such as ITU-R WP5A/5C/5D, IEEE 1900, ETSI, and CCSA.
\end{CCJNLbiography}

\begin{CCJNLbiography}{figures/Wen}{Wen Qin}
	received the M.S. degree from Xi’an University of Posts and Telecommunications, Xi’an, China, in 2022. He is currently pursuing the Ph.D. degree with the School of computer science, Northwestern Polytechnic University, Xi’an, China. His current research interests include the Avionics Intra-Communication Network and ad hoc network.
\end{CCJNLbiography}

\begin{CCJNLbiography}{figures/Zhou}{Ying Zhou}
	received the B.S. degree in communication engineering from Xidian University in 2019, M.S. degree in information and communication engineering from Beijing University of Information Science and Technology in 2022, and is currently studying for a Ph. D degree in information and communication engineering from Beijing University of Posts and Telecommunications. His research interests include wireless communications, resource allocation and vehicular networks 
\end{CCJNLbiography}

\begin{CCJNLbiography}{figures/Wei}{Zhiqing Wei}
	received his B.E. and Ph.D. degrees from BUPT in 2010 and 2015. Now he is an associate professor at BUPT. He was granted the Exemplary Reviewer of IEEE Wireless Communications Letters in 2017, the Best Paper Award of International Conference on Wireless Communications and Signal Processing 2018. He was the Registration Co-Chair of IEEE/CIC International Conference on Communications in China (ICCC) 2018 and the publication Co-Chair of IEEE/CIC ICCC 2019. His research interest is the performance analysis and optimization of mobile ad hoc networks.
\end{CCJNLbiography}

\begin{CCJNLbiography}{figures/Ping_Zhang}{Ping Zhang}
	(M'07-SM'15-F'18) received his M.S. degree in electrical engineering from Northwestern Polytechnical University, Xi'an, China, in 1986, and his Ph.D. degree in electric circuits and systems from BUPT, Beijing, China, in 1990. He is currently a Professor with BUPT. is currently a Professor with the School of Information and Communication Engineering, Beijing University of Posts and Telecommunications (BUPT), and the Director of the State Key Laboratory of Networking and Switching Technology. He is also an Academician with the Chinese Academy of Engineering (CAE). His research interests mainly focus on wireless communications. He is also a member of the IMT-2020 (5G) Experts Panel and the Experts Panel for China’s 6G development. He served as the Chief Scientist for the National Basic Research Program (973 Program), an Expert for the Information Technology Division of the National High-Tech Research and Development Program (863 Program), and a member of the Consultant Committee on International Cooperation, National Natural Science Foundation of China.

\end{CCJNLbiography}

\end{document}